\documentclass[twocolumn,useAMS,usenatbib]{mn2e}
\usepackage{graphicx}
\usepackage{natbib}

\usepackage{bm}
\usepackage{amsfonts}
\usepackage{latexsym}
\usepackage[latin1]{inputenc}
\usepackage{amsmath}
\usepackage{epsfig}
\usepackage{amsbsy}
\usepackage{amssymb}

\newcommand{\m}[1]{\boldsymbol{#1}}
\newcommand{\mtext}[1]{\hspace{0.6cm}\mbox{#1}\hspace{0.6cm}}

\newcommand{\mvb}[1]{\textbf{\emph{#1}}}

\newcommand{\n}{\mathrm n}
\newcommand{\p}{\mathrm p}
\newcommand{\e}{\mathrm e}
\newcommand{\bb}{\mathrm b}
\newcommand{\x}{\mathrm x}
\newcommand{\y}{\mathrm y}
\newcommand{\crt}{\mathrm c}

\def \nn  {\nonumber}

\newcommand{\bear}{\begin{eqnarray}}
\newcommand{\eear}{\end{eqnarray}}
\newcommand{\be}{\begin{equation}}
\newcommand{\ee}{\end{equation}}

\def\jnl@style{\it}
\def\aaref@jnl#1{{\jnl@style#1}}

\def\aaref@jnl#1{{\jnl@style#1}}

\def\aj{\aaref@jnl{AJ}}                   
\def\apj{\aaref@jnl{{\rm ApJ}}}                 
\def\apjl{\aaref@jnl{ApJ}}                
\def\apjs{\aaref@jnl{ApJS}}               
\def\apss{\aaref@jnl{Ap\&SS}}             
\def\aap{\aaref@jnl{{\rm A\&A}}}                
\def\aapr{\aaref@jnl{\rm A\&ARv}}          
\def\aaps{\aaref@jnl{A\&AS}}              
\def\mnras{\aaref@jnl{{\rm MNRAS}}}             
\def\prd{\aaref@jnl{\rm Phys.~Rev.~D}}        
\def\prl{\aaref@jnl{Phys.~Rev.~Lett.}}    
\def\qjras{\aaref@jnl{QJRAS}}             
\def\skytel{\aaref@jnl{S\&T}}             
\def\ssr{\aaref@jnl{Space~Sci.~Rev.}}     
\def\zap{\aaref@jnl{ZAp}}                 
\def\nat{\aaref@jnl{Nature}}              
\def\aplett{\aaref@jnl{Astrophys.~Lett.}} 
\def\apspr{\aaref@jnl{Astrophys.~Space~Phys.~Res.}} 
\def\physrep{\aaref@jnl{Phys.~Rep.}}      
\def\physscr{\aaref@jnl{Phys.~Scr}}       

\title[Ambipolar diffusion]{The relevance of ambipolar diffusion for neutron star evolution}

\author[A. Passamonti et al.]
{Andrea Passamonti$^{1,2}$\thanks{E-mail:passamonti@ua.es},  Taner Akg\"{u}n$^{1}$, Jos\'{e} A. Pons$^{1}$, Juan A. Miralles$^{1}$ 
\\ $^{1}$Department de Fisica Aplicada, Universitat d'Alacant, Ap. Correus 99, 03080 Alacant, Spain 
\\ $^{2}$INAF-Osservatorio Astronomico di Roma, via Frascati 44, I-00040, Monteporzio Catone (Roma), Italy}

\begin{document}

\date{\today}

\pagerange{\pageref{firstpage}--\pageref{lastpage}} \pubyear{}

\maketitle

\label{firstpage}


\begin{abstract}
We study ambipolar diffusion in strongly magnetised neutron stars,
with special focus on the effects of neutrino reaction rates and the
impact of a superfluid/superconducting transition in the neutron star
core. For axisymmetric magnetic field configurations, we determine the
deviation from $\beta-$equilibrium induced by the magnetic force and
calculate the velocity of the slow, quasi-stationary, ambipolar
drift. We study the temperature dependence of the velocity pattern and
clearly identify the transition to a predominantly solenoidal flow.
For stars without superconducting/superfluid constituents and with a
mixed poloidal-toroidal magnetic field of typical magnetar strength,
we find that ambipolar diffusion proceeds fast enough to have a
significant impact on the magnetic field evolution only at low core
temperatures, $T \lesssim 1-2 \times 10^8$~K. The ambipolar diffusion
timescale becomes appreciably shorter when fast neutrino reactions are
present, because the possibility to balance part of the magnetic force
with pressure gradients is reduced. We also find short ambipolar
diffusion timescales in the case of superconducting cores for $T
\lesssim 10^9$~K, due to the reduced interaction between protons and
neutrons. In the most favourable scenario, with fast neutrino
reactions and superconducting cores, ambipolar diffusion results in
advection velocities of several km/kyr. This velocity can
substantially reorganize magnetic fields in magnetar cores, in a way
that can only be confirmed by dynamical simulations.
\end{abstract}


\section{introduction} \label{sec:Intro}

The understanding of long term evolution of magnetic fields in neutron
stars (NSs) is crucial to connect possible evolutionary tracks between
NS classes.  The origin, structure and dynamics of the large-scale
magnetic fields in magnetars and high-B pulsars, and their influence
on observable emission processes have been the subject of many
studies.  The answers to these (strongly interrelated) issues must
explain why the magnetic field strength, inferred from astrophysical
observations, can vary by many orders of magnitude, from recycled
millisecond pulsars with dipolar magnetic fields of about
$10^{8}-10^{9}$G, to ``normal'' rotation powered pulsars with fields
between $10^{10}$ and $10^{13}$G, and superstrong fields of magnetars,
up to $10^{15}$G.  Besides their persistent emission, magnetars show
frequent outbursts and flares, which can release up to $10^{46}
\textrm{erg s}^{-1}$ \citep[for more details see
  e.g.][]{2008A&ARv..15..225M, 2015SSRv..191..315M,
  2015RPPh...78k6901T}. This rich phenomenology is usually attributed
to locally stronger fields (higher order multipoles, coronal loops,
strong crustal toroidal fields). Being isolated and slowly rotating
stars, neither accretion nor rotation can supply the required
energy. It is therefore important to understand the details of the
internal evolution of magnetic fields, and their links to the
magnetosphere and observable effects.

The internal magnetic field of a neutron star evolves mainly through
three processes: Ohmic diffusion, Hall drift and ambipolar diffusion
\citep{1992ApJ...395..250G, 1995MNRAS.273..643S}.  The combined effect
of Ohmic decay and Hall drift is dominant in the crust. Special
attention has been paid to the key role of the Hall drift in the crust
of NSs, with plenty of studies over the last decades 
\citep{HR2002,HR2004,Pons2007,Reise2007,Pons2009,Kondi2011,Vigano2012,Gourgo2013,Marchant,Gourgo2014,Gourgo2015a,Gourgo2015b}. Among
all, \cite{Vigano2013} performed the most complete study of the
magnetic and thermal evolution of isolated NSs, exploring the
influence of their initial magnetic field strength and geometry, their
mass, envelope composition and relevant microphysical parameters.
Using the same numerical code, Pons, Vigan\`{o} \& Rea (2013) showed
that a highly resistive layer in the deep crust is a crucial
ingredient for enhancing dissipation of magnetic energy of high-field
NSs. The majority of these works present 2D simulations, but recent 3D
simulations suggest that the Hall-induced small scale magnetic
features persist in the NS crust on longer time scales than in
axisymmetric 2D simulations, although the global evolution still tends
to the dipolar Hall attractor \citep{Wood2015}.

Concerning possible mechanisms operating in magnetar cores, the number
of works is sensibly smaller, and with far less detail than for the
crustal field evolution. Owing to its quadratic dependence on the
magnetic field strength, ambipolar diffusion was proposed
\citep{1992ApJ...395..250G} as the main process controlling the
evolution of magnetars during the first $10^3-10^5$ yr.  However, a
neutron star core cools down below the neutron-superfluid and
proton-superconducting critical temperatures very fast, and the
interaction between the various particle species and the magnetic
field becomes much more complex than in the standard MHD approach.
Nevertheless, \citet{1992ApJ...395..250G} argued that ambipolar
diffusion would still be a significant process in these cases.  On the
other hand, \citet{2011MNRAS.413.2021G} studied in more detail the
ambipolar diffusion in supefluid and superconducting stars, and
concluded that its role on the magnetic field evolution would be
negligible. This is one of the issues that we adress in this paper.
Other recent works \citep{Graber2015, Elfritz2016} have also shown
that, without considering ambipolar diffusion, the magnetic flux
expulsion from the NS core with superconducting protons is very slow.

The effect of ambipolar diffusion has been so far studied through
simple timescale estimates, with sparse isolated attempts to engage
simulations in a simplified 1D approach \citep{2008A&A...487..789H,
  2010MNRAS.408.1730H}.  In this work, we revisit this important topic
in a more detailed way, including realistic microphsyics inputs, with
the purpose of setting up the stage for multidimensional numerical
simulations. We aim at improving previous estimates by calculating
global velocity fields (as opposed to local estimates).  We will begin
by reviewing the theory, reconciling different notations and
assumptions, to obtain the equations describing the ambipolar
velocity. This includes an elliptical partial differential equation
which describes the local deviation from the chemical equilibrium due
to the magnetic force.  By solving numerically the velocity patterns
for given magnetic field topologies, we can identify in which NS
region ambipolar diffusion is more important, at which temperature,
and what the effect of a superconducting or superfluid phase
transition is.

The paper is organized as follows. In Section \ref{sec:Eq} we review
the formalism and the relevant equations. The magnetic field
configuration used in the calculations is described in Section
\ref{sec:mf}. Our numerical results for the ambipolar diffusion
velocity patterns obtained in different models are presented in
Section \ref{sec:res} and we discuss overall timescales in Section
\ref{sec:timescale}.  In Section \ref{sec:con} we summarize the main
conclusions and final remarks.

\section{Theoretical overview} \label{sec:Eq}

Ambipolar diffusion is a mechanism only present in multicomponent
systems. In their seminal work \cite{1992ApJ...395..250G} start from
the equations of motion of charged particles in the presence of a
magnetic field and a fixed background of neutrons to derive the
relevant equations, under a number of simplifying assumptions. A more
general description of multifluid magnetohydrodynamics, including the
discussion of superfluid and superconducting components is the work of
\cite{2011MNRAS.410..805G}. The latter reference describes a rigorous
covariant formalism to treat general multifluid problems. In our case,
we consider a magnetised fluid made of three particle species,
protons, electrons and neutrons, respectively.  We assume they
interact through scattering processes mediated by electromagnetic
(charged particles) and nuclear forces (between protons and neutrons),
and are subject to $\beta$-reactions (weak interactions), and to the
gravitational potential.

We are interested in the quasi-stationary evolution driven by slow
motions, on timescales much longer than all the relaxation times
between collisions of particles of different species. Therefore, we
can safely neglect inertial terms in the equations of momentum, as
well as terms of order $v_i^2$.

We begin by considering the case of a non-rotating neutron star
composed of normal (non-superconducting and non-superfluid) matter.
To identify the fluid constituents in this work we use Roman letters
(x, y).  Specifically, we denote neutrons, protons, and electrons
with the letters $\n$, $\p$ and $\e$, respectively.  The three force balance
equations, one for each particle species are:
\begin{eqnarray}
 - \m{\nabla} \mu_{\p} - m_{\p}^{*}
\m{\nabla} \Phi + e \left( \mvb{E} + \frac{\mvb{v}_{\p}}{c} \times \mvb{B} \right) \!\!\!&=\!\!& \frac{m_{\p}^{*} \mvb{w}_{\p\n}}{\tau_{\p \n}} + \frac{m_{\p}^* \mvb{w}_{\p\e}}{\tau_{\p \e}} \nonumber \\ 
-\m{\nabla} \mu_{\e} - m_{\e}^{*}  \m{\nabla} \Phi - e \left( \mvb{E} + \frac{\mvb{v}_{\e}}{c} \times \mvb{B} \right) \!\!&=\!\!& \frac{m_{\e}^{*}  \mvb{w}_{\e\n}}{\tau_{\e \n}} + \frac{m_{\e}^{*} \mvb{w}_{\e\p}}{\tau_ {\e \p}} \nonumber \\ 
  - \m{\nabla} \mu_{\n} - m_{\n}^{*} \m{\nabla} \Phi \!\!&=\!\!&
\frac{m_{\n}^{*} \mvb{w}_{\n\p}}{\tau_{\n \p}} + \frac{m_{\n}^{*}
  \mvb{w}_{\n\e}}{\tau_{\n \e}} \nonumber \\
 \label{3eqs}
\end{eqnarray}
where $\mvb{E}$ is the electric field, $\mvb{B}$ is the magnetic field
$\mvb{w}_{\x\y} = \mvb{v}_{\x} - \mvb{v}_{\y}$ are the relative
velocities between the different ``fluids'', $\Phi$ is the
gravitational potential, $\mu_{\x}$ the chemical potentials,
$m_{\x}^{*}$ the effective masses, and $\tau_{\x\y}$ is the relaxation
time for collisions of $\x-$particles with $\y-$particles.  In
general, electrons are degenerate relativistic particles in the core
of neutron stars, and their effective mass can be considerably larger
than their rest mass, $m_e^*=m_e (1+x_e^2)^{1/2}$, with $x_e$ being
the ratio between the Fermi momentum and the electron rest mass
$x_e\equiv p_F/m_e c$. On the contrary, the effective masses of
neutrons and protons (non-relativistic) contain an interaction
contribution that results in effective masses typically smaller than
the rest masses $m_n^*/m_n \approx m_p^*/m_p \approx 0.6-0.8$. In this
work we consider constant effective masses for protons and neutrons
with $m_p^{*} = 0.6 \, m_p$ and $m_n^{*} = 0.75 \, m_n$.

Conservation of momentum implies the conditions ${n_{\x}
  m_{\x}^*}/{\tau_{\x \y}}= {n_{\y} m_{\y}^{*}}/{\tau_{\x \y}}$, with
$n_{\x}$ denoting the number density of $\x-$type particles.  We use
the simple description of friction in terms of relaxation times,
following \cite{1992ApJ...395..250G}, but equivalent expressions can
be derived from the more formal equations of
\cite{2011MNRAS.410..805G}, which use {\it entrainment} coefficients
to model the coupling between different species.

Combining all three equations to remove the collision terms, one
arrives at 
\begin{equation} 
n_{\rm c} \m{\nabla} (\Delta \mu) + n_{\rm b} {\m{\nabla}} \mu_{\n} +
\rho {\m{\nabla}} \Phi = \frac{\mvb{j} \times \mvb{B} }{c}~,
\label{mhdeq}
\end{equation}
 where we have used the local charge neutrality $n_{\rm c} \equiv
 n_{\e} \approx n_{\p}$, while $n_{\rm b}=n_{\p}+n_{\n}$ is the baryon
 number density, $\rho= m_{\p}^{*} n_{\p} + m_{\n}^{*} n_{\n} +
 m_{\e}^{*} n_{\e}$ represents the total mass density, $\mvb{j} = e
 n_{\rm c} \mvb{w}_{\p\e}$ is the electric current density, and
 $\Delta \mu \equiv \mu_{\p}+\mu_{\e}-\mu_{\n}$ is the deviation from
 $\beta$-equilibrium. The right hand side is the magnetic force acting
 on the fluid, which for a non-superconducting star is given by the
 Lorentz force:
\begin{equation}  
\mvb{f}_{\!  mag} \equiv \frac {\mvb{j} \times \mvb{B}}{c} \, .
\end{equation}
 In a strict magnetostatic equilibrium $\mvb{f}_{\!  mag}$ must
 exactly balance the left hand side of equation (\ref{mhdeq}).

The formal derivation of the evolution equations proceeds as
follows. First, by combining Eqs. (\ref{3eqs}), one can work out a
general expression of the electric field in terms of the electric
current, plus additional terms (generalized Ohm's law), one of which
involves a relative velocity between two species. This electric field
enters the induction equation describing the evolution of the magnetic
field: \be \frac{\partial \mvb{B}}{\partial t} = - c \m{\nabla} \times
\mvb{E} \, . \ee

In order not to carry on unnecessary coefficients, at this point we
note that, for typical neutron star conditions, proton-neutron
scattering is mediated by the strong force, while the electrons
interact only electromagnetically with the weak neutron magnetic
moment.  Therefore, we can safely assume that
\begin{equation}
\frac{m_\e^{\ast}  }{\tau_{\e \n} } \ll \frac{m_\p^*  }{\tau_{\p \n} }  \, . \label{eq:cond}
\end{equation}
For simplicity, we also neglect the electron mass (electron
contributions to gravitational terms and total momentum). 

Then, from the electron momentum equation and the definition of the current, we
obtain the following expression for the electric field:
\begin{equation}
\mvb{E} = \frac{\mvb{j}}{\sigma_0} - \frac{1}{c} \mvb{v}_{\p} \times \mvb{B} +
\frac{1}{e n_{\rm c} c} ~ \mvb{j} \times \mvb{B} - \frac{1}{e} \m{\nabla}
\mu_e \, ,
\label{tot2}
\end{equation}
where $\sigma_0 = e^2 n_{\rm c} \tau_{\e \p}/m_\e^*$ is
the electrical conductivity in the absence of a magnetic field, which,
in the region of validity of equation (\ref{eq:cond}), is dominated by
electron-proton collisions. We note that the last term in
equation (\ref{tot2}) is irrelevant for the induction equation, since its
curl vanishes.  The first term on the right-hand side is the Ohmic
dissipation and the third term is the Hall term, both of which are
very important in the NS crust, but are negligible in the core.

If we define the average baryon velocity by
\begin{equation}
\mvb{v}_{\rm b} = \frac{n_\n  \mvb{v}_{\n} + n_\p  \mvb{v}_{\p} }{n_\bb} \, , 
\end{equation}
we can rewrite the second term on the right hand side as
\begin{equation}
\mvb{v}_{\p} \times \mvb{B} = \mvb{v}_{\bb} \times \mvb{B} + x_\n \mvb{w}_{\p \n} \times \mvb{B} \, , 
\end{equation}
where $x_{\n} = n_{\n} / n_{\rm b}$ is the neutron fraction.  In this
form, we can identify the advective term due to the hydrodynamic
velocity of baryons, which should be negligible if we are very close
to dynamical equilibrium, and the second term, the {\it ambipolar
  diffusion}, due to relative velocity between protons and neutrons.
Note that ambipolar diffusion becomes dominant over the Hall term when
protons and electrons are strongly coupled, with their velocity
difference being much smaller than their individual
velocities. Therefore, our problem can be reduced to a two-fluid model
consisting of a neutral component and a charged fluid (protons plus
electrons, moving with nearly the same speed) locked to the magnetic
field. If the charged and neutral components are also locked to each
other, there is only a single hydrodynamic velocity and we recover the
one-fluid MHD limit.
\subsection{Ambipolar drift velocity} \label{sec:amb}

 We now discuss how to estimate the ambipolar diffusion velocity,
 $\mvb{v}_{amb} \equiv x_\n \mvb{w}_{\p \n}$.  Combining the three
 equations (\ref{3eqs}) to remove the electric field and using
 equation~(\ref{eq:cond}), we have:
\begin{equation}
  \frac{ \mvb{f}_{\! mag} }{ n_{\crt}}   -\m{\nabla} \left( \Delta \mu \right)  
  = \frac{1}{x_\n^2} \frac{m_\p^* \, \mvb{v}_{amb} }{\tau_{\p \n} } 
 \label{eq:vamb} \, .
\end{equation}

If $\beta$-reactions are fast, bringing the fluid to chemical equilibrium
($\Delta \mu=0$), Equation (\ref{eq:vamb}) shows that there is a
quasi-stationary, slow motion of the charged component with respect to
the neutron fluid, proportional to the Lorentz force, simply
\begin{equation}
 \mvb{v}_{amb}  = x_\n^2  \frac{\tau_{\p \n} }{m_\p^* \, }  \frac{ \mvb{f}_{\! mag} }{ n_{\crt}}  ~. 
\label{eq:vhigh}
\end{equation}

Equation (\ref{eq:vamb}) also shows that, if $\beta$-equilibrium is
not reached faster than the evolution timescale, the magnetic force
per charged particle can be partially balanced by the pressure
gradients induced by small deviations from
$\beta$-equilibrium. However, we note that only the irrotational part
of $\mvb{f}_{\! mag} / n_{\crt}$ can be cancelled by a gradient term,
the solenoidal part remains unbalanced resulting in a finite ambipolar
velocity (see Appendix \ref{sec:app1} for more details on the
irrotational-solenoidal decomposition of a vector field).

In order to determine the ambipolar velocity in the general case, we need to calculate the chemical
deviation $\Delta \mu$ throughout the star. For this purpose, we must also consider
the individual continuity equations and, since we search for quasi-stationary solutions, we can
neglect the time variation of the number densities to write:
\begin{eqnarray}
 \m{\nabla} \cdot \left( n_{\p}
\mvb{v}_{\p} \right)&=& - \Delta \Gamma ~, \nonumber \\ 
 \m{\nabla} \cdot \left( n_{\e}
\mvb{v}_{\e} \right) &=& - \Delta \Gamma ~, \nonumber \\
\m{\nabla} \cdot \left( n_{\n}
\mvb{v}_{\n} \right) &=& \Delta \Gamma .
\label{eq:con1}
\end{eqnarray}
where 
\begin{equation}
\Delta \Gamma = \Gamma \left( p + e \to n +  \nu_{\e} \right) - \Gamma \left(n \to p + e + \bar{\nu}_{\e} \right) \, ,  
\end{equation}
with $\Gamma$ denoting the reaction rate.
When $\Delta \mu \ll k_{B} T$, the reaction rates can be linearized
and written in terms of the deviation from chemical equilibrium as
follows:
\begin{equation}
\Delta \Gamma = \lambda \Delta \mu \, ,  \label{eq:dG}
\end{equation}
where  $\lambda \equiv \left(
\textrm{d} \Gamma / \textrm{d}\Delta\mu \right)|_{eq}$ is a
coefficient which depends on the density and temperature.  
If $\Delta \mu \gtrsim k_{B} T $, nonlinear terms in the
$\beta$-reaction rates should be considered.

\begin{figure*}
\begin{center}
\includegraphics[height=80mm]{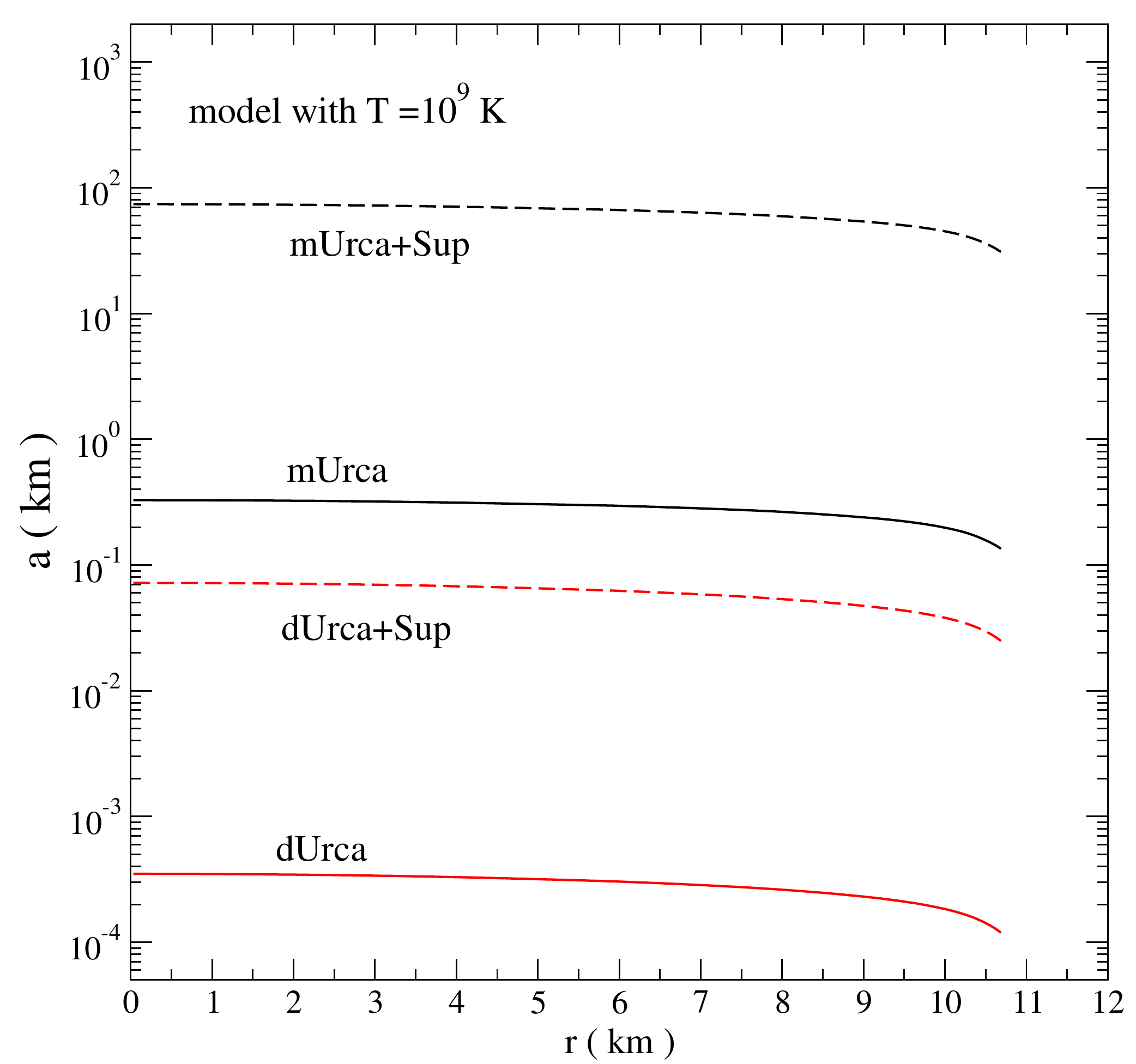}
\includegraphics[height=80mm]{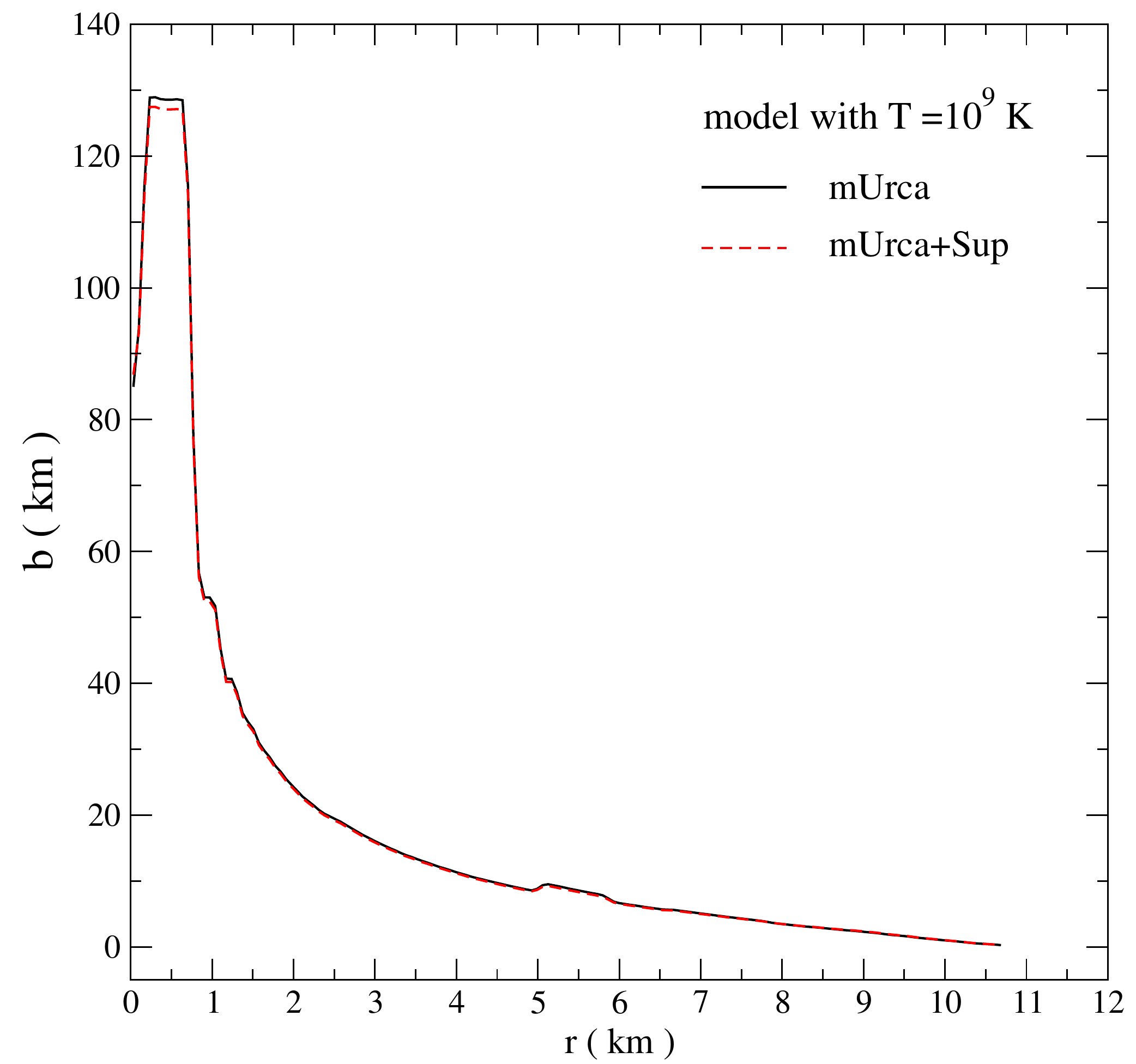}
\caption{Radial profiles of the coefficients $a$ and $b$ (see
  Eq. \ref{eq:ab}) for a star with $T = 10^9$~K. The left-hand panel
  displays the coefficient $a$, on logarithmic scale, for stellar
  models with normal (solid lines) and superfluid/superconducting
  matter (dashed lines) and for mUrca (black lines) and dUrca
  reactions (red lines). The right-hand panel shows the coefficient
  $b$ for a star with mUrca reactions and with normal (solid-black
  line) and superfluid/superconducting (dashed-red line) matter. In
  this figure the superconducting model has $T_{c\rm p}=5\times10^9$~K
  and $T_{c\rm n}=10^9$~K.
\label{fig:ab}}
\end{center}
\end{figure*}

Taking the divergence of Eq. (\ref{eq:vamb}) and making use of the
continuity Eqs. (\ref{eq:con1}), we can obtain the following elliptic
equation for the chemical equilibrium deviation $\Delta \mu$:
\begin{equation}
 \nabla^2  \left( \Delta \mu \right) 
-  \frac{1 }{b}   \,    \frac{ \partial \Delta \mu }{\partial r} 
-  \frac{1}{a^2}  \, \Delta \mu
=  \m{\nabla} \cdot  \left( \frac{ \mvb{f}_{\! mag} }{ n_{\crt}}      \right)
- \frac{1 }{b}  \, \frac{ f^{r}_{ mag} }{ n_{\crt}}    
\, ,   \label{eq:ell}
\end{equation}
where  we have defined 
\begin{align}
& \frac{1}{a^2} = \frac{\lambda \, m_{\p}^* }{x_{\n}^2 n_{\crt} \tau_{\p \n} } \, , \qquad \qquad  
 \frac{1}{b} =  \frac{d}{d r} \ln \left(\frac{m_{\p}^*}{x_{\n} n_{\crt} \tau_{\p \n}}   \right) \, .
\label{eq:ab}
\end{align}
Both $a$ and $b$ have dimensions of length.  In Appendix
\ref{sec:app0} we give more details about the derivation of equation
(\ref{eq:ell}) and the approximations involved.  To summarize, for a
given magnetic field configuration we can calculate $\mvb{v}_{amb}$
from equation (\ref{eq:vamb}), where $\Delta \mu$ is obtained by
solving numerically equation (\ref{eq:ell}) in the domain $ 0\leq r
\leq R_{cc}$ and $ 0\leq \theta \leq \pi$, where $R_{cc}$ denotes the
position of the crust/core interface. As boundary conditions, we
impose that the ambipolar velocity satisfies the regularity conditions
at the origin and the magnetic axis.  At the crust-core interface,
($r=R_{cc}$), we impose that the radial component of the velocity
vanishes, which is equivalent to:
\begin{equation}
  \frac{ \partial \Delta \mu }{\partial r }- \frac{ {f}^r_{ mag} }{n_{\rm c}} =0  \, .
\end{equation}
This boundary condition is easily satisfied when $\Delta \mu$ is not
negligible.  On the other hand, at high temperatures, $T\sim 10^9$K,
$\beta$-reactions quickly smooth out deviations from equilibrium and a
discontinuity (or a sharp gradient) in $\Delta \mu$ is developed at
the crust-core interface, if the magnetic force does not vanish there.
Some numerical tests of the solutions against analytical models are
presented in Appendix \ref{sec:app1}.

As for the microphysics input, for normal matter, we adopt the
proton-neutron collision time, $\tau_{\p\n}$,
from~\cite{1990SvAL...16...86Y}:
\begin{equation}
 \frac{1}{\tau_{\p\n} } = 4.7 \times 10^{16} T_8 ^2 \left( \frac{ \rho }{\rho_{nuc}} \right) ^{-1/3}  \textrm{s}^{-1} \, , 
 \label{eq:taupn}
\end{equation}
where $T_8$ is the temperature in units of $10^{8}$K, and $\rho_{nuc}
= 2.8 \times10^{14} \textrm{g cm}^{-3}$ is the nuclear saturation
density.

In the neutron star interior, weak interactions are driven by two
types of Urca reactions. The most common channel is the modified Urca
(mUrca), which allows the conservation of energy and momentum with the
help of a spectator nucleon. In very compact stars, if the proton
fraction $x_{\p} \gtrsim 0.11$, $\beta$-reactions could be much faster
through the activation of direct Urca (dUrca) processes, without the
support of spectator nucleons \citep{1991PhRvL..66.2701L}.
We will use the mUrca reaction rates given by
\cite{1989PhRvD..39.3804S}:
\begin{equation}
 \lambda  = 5 \times 10^{27} T_8 ^6 \left( \frac{ \rho }{\rho_{nuc}} \right) ^{2/3} 
 \textrm{ergs}^{-1}  \textrm{cm}^{-3}  \textrm{s}^{-1}  \, ,  \label{eq:LM}
\end{equation}
and the dUrca rates obtained by \cite{1992PhRvD..45.4708H}
\begin{equation}
 \lambda  = 3.5 \times 10^{36}   \frac{m_{\n}^{\ast}}{m_\n} \frac{m_{\p}^{\ast}}{m_\p}
 T_8 ^4 \left( \frac{ \rho_{\crt} }{\rho_{nuc}} \right) ^{1/3} 
  \textrm{ergs}^{-1}  \textrm{cm}^{-3}  \textrm{s}^{-1}  \, . \label{eq:LD}
\end{equation}

\subsection{Superfluidity and Superconductivity} \label{sec:sup0}

After a neutron star cools down below the critical temperatures
$T_{c\n}$ and $T_{c\p}$, the neutrons and protons in the core are,
respectively, in a superfluid and superconducting state. In a type II
superconductor, the magnetic field permeates the core with an array of
quantised fluxtubes, while neutron vortices sustain the star's
rotation.  Collision rates and $\beta$-reaction rates decrease
considerably with respect to the normal matter case, and new
dissipative processes mediated by vortices may appear.  Since we are
interested in strong magnetic fields (and therefore slowly rotating
stars), we neglect the effects of neutron vortices on the long term
evolution.  In other words, we are in the hot superfluid regime
studied by \citet{2011MNRAS.413.2021G} ($ T \gtrsim 3 \times 10^8$K)
in which particle scattering is dominant over the dissipation due to
mutual friction between fluxoids and vortices.

Under this assumption, we introduce the superfluidity and
superconductivity effects in our analysis by considering the
corrections to both the relaxation timescale $\tau_{\p\n}$ and the
weak reaction rate $\lambda$, which are usually expressed in terms of
suppression factors, respectively $\mathcal{R}_{\p\n}$ and
$\mathcal{R}_{sup}$, which become exponentially small in the strong
superfluid regime $T \ll T_{c\x}$. In general, the suppression factors
depend on the gap model and are described by complex integrals, whose
results have been fitted to more practical analytical formulae
\citep{1994AstL...20...43L, 1995A&A...297..717Y, 2000A&A...357.1157H,
  2001A&A...372..130H}.  From these references, we take the
appropriate suppression factors to modify the rates as follows:
 \begin{align}
  \frac{1}{\tau_{\p\n} }& \longrightarrow   \frac{\mathcal{R}_{\p\n}}{\tau_{\p\n} }   \, , \label{eq:TR} \\ 
  \lambda  & \longrightarrow   \lambda \, \mathcal{R}_{sup} \, , \label{eq:LR}
\end{align} 
and consequently the coefficients of Eqs.~(\ref{eq:vamb}) and
(\ref{eq:ell}).

We also consider the superconductivity effects on the force by
replacing the Lorentz force with the magnetic force for a type
II superconductor \citep{2008MNRAS.383.1551A, 2011MNRAS.410..805G, 2011MNRAS.413.2021G}:
\begin{equation}
\mvb{f}_{\! mag} = \frac{1}{4\pi} \left( \m{\nabla} \times H_{c1} \m{\hat{B}} \right)
\times \mvb{B} - \frac{ n_{\rm c} }{4 \pi} \m{\nabla} \left( B
\frac{\partial H_{c1}}{\partial n_{\rm p}} \right)  \, ,
\end{equation}
where $\m{\hat{B}} = \mvb{B}/B$ is the unit vector in the direction of the
magnetic field, and $H_{c1}$ is the lower critical field
\citep{Tinkham2004}. In typical conditions of type II superconductivity in NSs
\begin{equation}
H_{c1} \approx 10^{15} \left(\frac{n_\p}{0.01 \,  \textrm{fm}^{-3}}\right)~{\mathrm G}.
\end{equation}

The radial profiles of the coefficients $a$ and $b$ defined in
equation~(\ref{eq:ab}) are shown in Fig. \ref{fig:ab} for different
states of matter (normal and superfluid/superconducting) and weak
reaction processes (mUrca and dUrca). The stellar model is built with
the same equation of state as in \cite{Vigano2013}, and its parameters
are $M=1.4 M_\odot$, $R=11.6$~km, while the crust/core interface is at
$R_{cc}=10.79$~km.  As an illustrative example we show the results for
a star with $T=10^9$~K and with a constant gap model described by
$T_{c\p} = 5\times 10^9$~K and $T_{c\n} = 10^9$~K.  The left-hand
panel of Fig.  \ref{fig:ab} shows the strong dependence of $a$ on the
$\beta$-reaction process. Its value for mUrca reactions is about three
orders of magnitude larger than the dUrca case. In the same figure we
can notice the effects of the superconducting transition and thus of
the suppression factors $\mathcal{R}_{\p\n}$ and $\mathcal{R}_{sup}$,
which increase $a$ for more than two orders of magnitude. When $T \ll
T_{c\p}$, the effects of the suppression factors are even more
relevant due to their exponential dependence on the temperature.  The
quantity $b$ does not depend on $\lambda$ and thus it is not affected
by the particular $\beta$-reaction process.  Hence, we show in the
right-hand panel of Fig. \ref{fig:ab} the profile of $b$ for the mUrca
reactions and for normal and superconducting matter.  Since
$\mathcal{R}_{\p\n}$ is almost constant for the gap model used in this
work, the $\mathcal{R}_{\p\n}$ has a tiny effect on $b$.  For stars in
a non superfluid/superconducting state, the variation of $a$ and $b$
with the temperature can be easily determined by
equations~(\ref{eq:ab}) and (\ref{eq:taupn})-(\ref{eq:LD}).

\section{Magnetic field configuration}  \label{sec:mf}

The actual geometry of the magnetic field inside a neutron star is
unknown. For practical purposes, we will consider an analytical,
axisymmetric model which satisfies the relevant boundary and
regularity conditions.

Any axisymmetric magnetic field can be decomposed into poloidal and toroidal components as follows
\citep{1961hhs..book.....C}
	\begin{equation}
	\m{B} = \frac{1}{r \sin \theta } \left(  \m{\nabla} {\cal P} \times \m{ \hat \phi}  +  {\cal T} \m{\hat{\phi}} \right) \, .
	\end{equation}
Here ${\cal P}(r,\theta)$ and ${\cal T}(r,\theta)$ are, respectively,
the poloidal and toroidal stream functions, and $\m{ \hat \phi}$ is
the unit vector in the azimuthal direction. In a barotropic fluid in
MHD equilibrium, these functions must be solutions of the
Grad--Shafranov equation, and the Lorentz force can be expressed as
the mass density times the gradient of a (magnetic) potential,
$\m{f}_{mag} = \rho \m\nabla {\cal M}({\cal P})$, where ${\cal M}$ is
some arbitrary function of ${\cal P}$. In this case $\m{f}_{mag} /
\rho$ is a purely irrotational quantity.  On the other hand, in a
non-barotropic star the quantity $\m{f}_{mag} / \rho$ is not necessarily
a gradient of a potential, and the poloidal and toroidal functions can be chosen
with more flexibility.  Thus, we adopt the simple magnetic field
model constructed for non-barotropic fluids in
\cite{2013MNRAS.433.2445A}.

We consider a mixed poloidal-toroidal configuration which smoothly
joins to a vacuum dipole solution at the star's surface. We choose a
dipolar poloidal function of the form
\begin{equation}
{\cal P}(r,\theta) = {\cal P}_{0} f(x)\sin^2\theta \ , \label{eq:pfunc}
\end{equation}
where ${\cal P}_0$ is a constant that sets the poloidal field
amplitude, and $x$ is a dimensionless radial coordinate defined
through $x=r/R_{\star}$, with $R_{\star}$ being the radius where the
field continuously joins the vacuum solution. In this work,
$R_{\star}$ is taken as the stellar radius.
\begin{figure}
\begin{center}
\includegraphics[trim = 42mm 12mm 38mm 10mm, clip,  height=59mm]{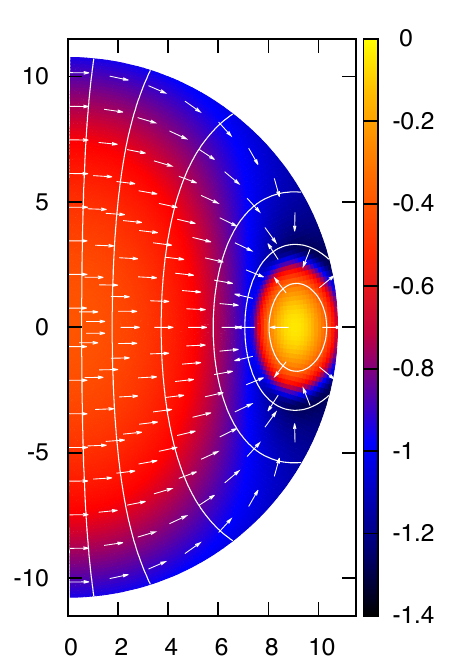}
\includegraphics[trim = 42mm 12mm 38mm 10mm, clip, height=59mm]{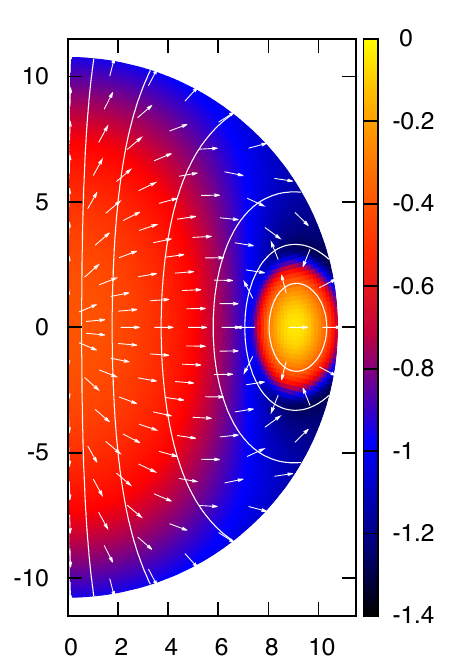}
\caption{Configuration of the magnetic field and the magnetic force.
  The colors denote the strength of the magnetic field magnitude in
  logarithmic scale, $\log \left[B/\left(10^{15} G\right) \right] $,
  the white-solid curves show the projection of the poloidal magnetic
  field lines, while the arrows represent the direction of the
  magnetic force.  In the two panels, the magnetic field has the same
  mixed poloidal-toroidal geometry with ${\cal P}_{\rm c} = {\cal
    P}_0$.  The magnetic force is the usual Lorentz force (left-hand
  panel) and superconducting force (right-hand panel).  In the
  horizontal and vertical axes (left), the units are given in km.
\label{fig:Bconf}}
\end{center}
\end{figure}
Outside the star, where there are no currents, the dipole vacuum
solution must satisfy
\begin{equation}
f_{\rm out}(x) \propto x^{-1}  \, . \label{eq:fout}
\end{equation}
On the other hand, the interior dipolar field is assumed to be a
polynomial of order $n$
\begin{equation}
f_{\rm in}(x) = \sum_{i=1}^n f_i x^i  \, ,  \label{eq:fin}
\end{equation}
where $f_i$ are coefficients to be determined from regularity
conditions at the center and boundary conditions at the surface. These
conditions imply that the function $f_{\rm in}$ must have at least
three terms; in particular, we take the first three even terms in the
power series \citep[see][for a detailed discussion of the regularity
  and boundary conditions]{2013MNRAS.433.2445A}. Thus, the radial
function can be written as
	\begin{equation}
	f(x) = \left\{
	\begin{aligned}
	& f_2 x^2 + f_4 x^4 + f_6 x^6 & \mtext{for} x < 1 \, , \\
	&  x^{-1} & \mtext{for} x \geqslant 1 \, ,
	\end{aligned}
	\right.         \label{eq:fp}
	\end{equation}
where the vacuum solution is normalized so that $f(1) = 1$. The three
unknown coefficients are determined from the boundary conditions at
$x=1$ for the continuity of the magnetic field and the vanishing of
the current, which can now be expressed as
	\begin{equation}
	f'(1) = - f(1)
	\mtext{and}  f''(1) =  2 f(1) \ .  \label{eq:bdm}
	\end{equation}
The coefficients which satisfy these boundary conditions are given by
	\begin{equation}
	f_2 = \frac{35}{8} \ ,
	\hspace{0.5cm} f_4 = - \frac{21}{4} 
	\mtext{and} f_6 = \frac{15}{8}   \, .
	\end{equation}

We take the toroidal component of the magnetic field to be described
by a toroidal function ${\cal T}({\cal P})$ which is confined to the
closed field lines within the radius $x = 1$. This assumption implies
that the azimuthal component of the Lorentz force vanishes. Following
the notation of \citet{2016arXiv160502253A}, we choose
	\begin{equation}
	{\cal T}({\cal P}) = \left\{
	\begin{aligned}
	& s({\cal P} - {\cal P}_{\rm c})^\sigma & \mtext{for} {\cal P} \geqslant {\cal P}_{\rm c} \ , \\
	& 0 & \mtext{for} {\cal P} < {\cal P}_{\rm c} \ .
	\end{aligned}
	\right.
	\end{equation}
where $s$ is a constant that sets the amplitude of the toroidal field
with respect to the poloidal field, $\sigma$ is a constant that
defines the relation between the functions ${\cal T}$ and ${\cal P}$,
and ${\cal P}_{\rm c}$ defines the field line which encloses the
toroidal field. In this work, we choose ${\cal P}_{\rm c}$ to be equal
to the maximum value of the function ${\cal P}$ at the stellar
surface, i.e. ${\cal P}_{\rm c} = {\cal P}_0$. In order to avoid
surface currents we must have $\sigma \ge 1$. Moreover, to ensure the
continuity of the gradient of the Lorentz force across the toroidal
boundary, we set $\sigma=2$.

As described in Sec.~\ref{sec:sup0}, when the star is superconducting
we replace the Lorentz force with the superconducting magnetic force.
Nevertheless, we still use the same magnetic field configuration
described in this section. Although this approach is not strictly
correct, for the purposes of this work we prefer to maintain here the
same magnetic field configuration in order to isolate the effects of a
different superconducting force and the reduced collision rates on the
velocity pattern.  A further effect related to the change of magnetic
topology in models with superconducting cores
\citep{Roberts1981,HW2013,2013PhRvL.110g1101L,2014MNRAS.437..424L}
will be addressed in the future. In this work, our purpose is to give
quantitative estimates for the different possible scenarios.

For a mixed poloidal-toroidal configuration, with $10^{14}$~G at the
pole and a toroidal field with maximum strength $10^{15}$~G, we show
in Fig. \ref{fig:Bconf} the magnetic field amplitude and its field
lines as well as the direction of the Lorentz (left-hand panel) and
superconducting (right-hand panel) forces.  Obviously, the choice of
the initial configuration determines the value of the Lorentz force
and therefore $\mvb{v}_{amb}$.
\begin{figure}
\begin{center}
\includegraphics[trim = 42mm 12mm 38mm 10mm, clip, height=59mm]{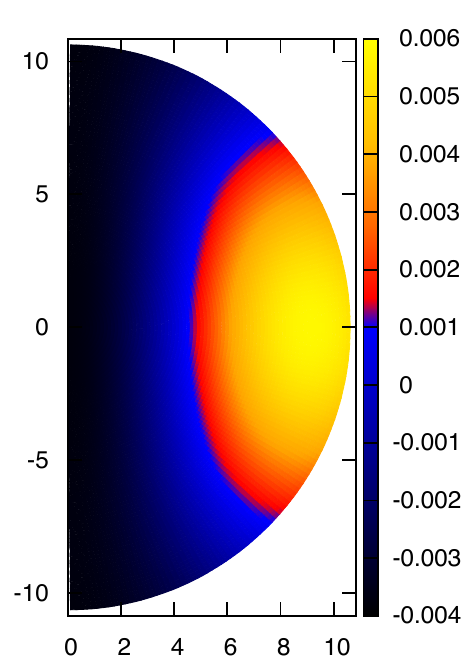}
\includegraphics[trim = 42mm 12mm 38mm 10mm, clip, height=59mm]{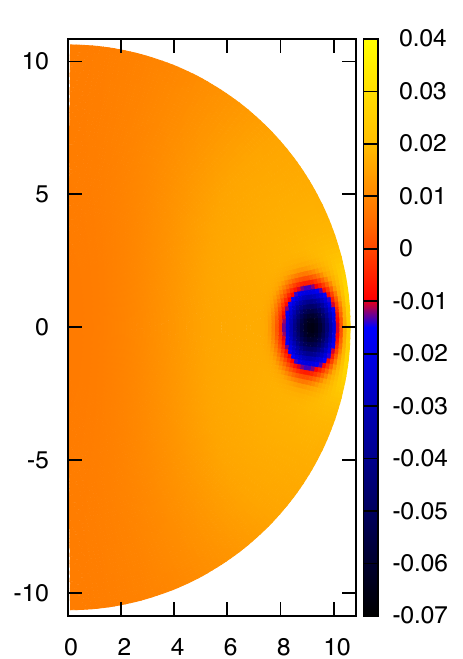}
\caption{Two-dimensional contour plots of the quantity $\Delta \mu /
  k_{\rm B} T$ for a normal matter star, with mUrca neutrino
  processes, and $T=2\times 10^{8}$K. The left and right-hand panels
  show, respectively, the solution for a purely poloidal field (model
  A) and for a mixed poloidal-toroidal field (model B). In the
  horizontal and vertical axes (left), the units are given in km.
\label{fig:Dmu}}
\end{center}
\end{figure}

\begin{figure*}
\begin{center}
\includegraphics[trim = 42mm 12mm 38mm 10mm, clip, height=58mm]{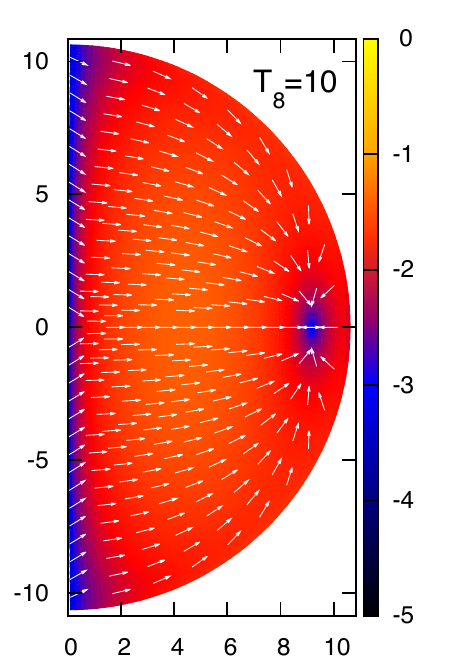}
\includegraphics[trim = 42mm 12mm 38mm 10mm, clip, height=58mm]{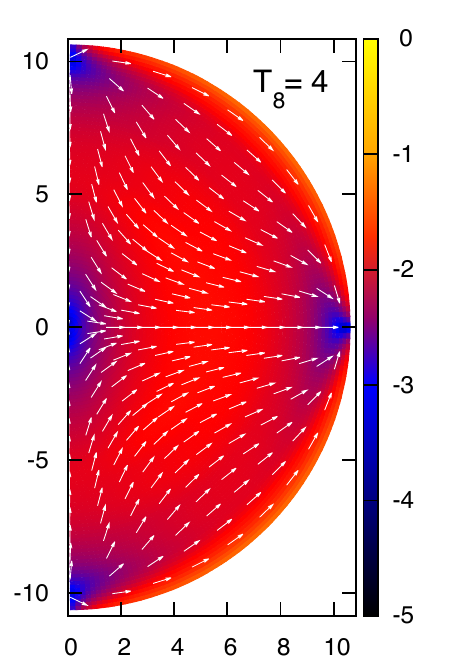}  
\includegraphics[trim = 42mm 12mm 38mm 10mm, clip, height=58mm]{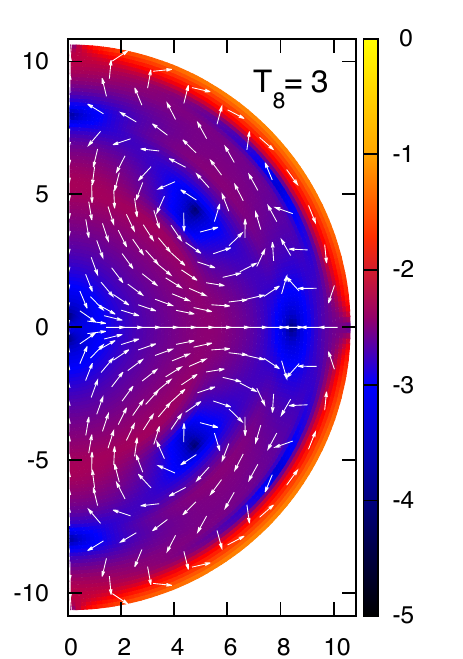} 
\includegraphics[trim = 42mm 12mm 38mm 10mm, clip, height=58mm]{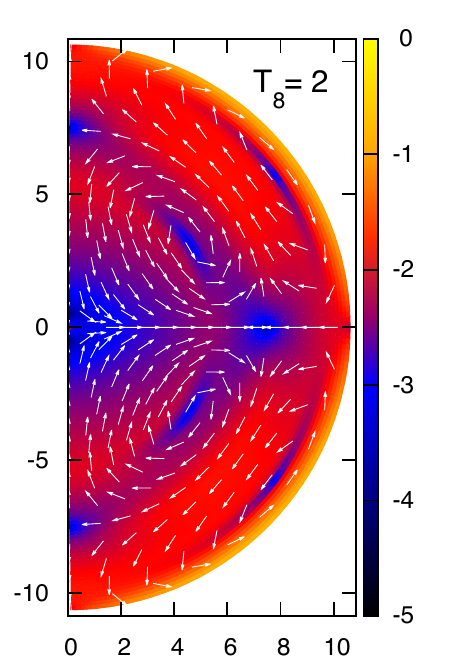} 
\caption{Ambipolar diffusion velocity ($\mvb{v}_{amb}$) in a
  non-superfluid/superconducting core with mUrca processes at
  different temperatures. From the left to right: $T_{8} = 10, 4, 3$
  and $2$.  The magnetic field is given by model A (purely poloidal
  field).  The arrows show the direction of the velocity field and the
  color scale represents $\log|\mvb{v}_{amb}|$, where velocities are
  given in km/Myr. In the horizontal and vertical axes (left), the
  units are given in km.
\label{fig:Vamb-MUrca}}
\end{center}
\end{figure*}

\section{Results} \label{sec:res}

In this section, we present our results for the axisymmetric magnetic
field configuration described in Sec. \ref{sec:mf}.  The neutron star
model is built with the same equation of state and parameters as in
\cite{Vigano2013} ($M=1.4 M_\odot$, $R=11.6$ km and $R_{cc}=10.79$ km).  We consider two
magnetic field models: i) a purely poloidal magnetic field with $B_{p}
= 10^{14}$G at the pole (hereafter model A); ii) a mixed
poloidal-toroidal magnetic field with $B_{p} = 10^{14}$G and a
toroidal field with maximum strength $B_{t} = 10^{15}$G (hereafter
model B).  We explore the temperature interval $10^{7}\textrm{K} \leq
T \leq 2\times 10^{9}$K, which covers the expected core temperatures
in a NS from $1-10^6$ years. Note that the core becomes nearly
isothermal (except for gravitational redshift corrections) only
minutes after birth, and that important thermal gradients are only
present in the envelope and to a lesser extent in the crust, in
presence of strong magnetic fields. We also remind that this internal
temperature is not the surface temperature, which is typically two orders of
magnitude smaller.  We consider both normal and
superfluid/superconducting matter and discuss the differences between
the standard cooling scenario (mUrca reactions) and the fast cooling
scenario (dUrca processes).

To derive equation (\ref{eq:ell}) we have approximated the
$\beta$-reaction rate with equation (\ref{eq:dG}). However, this
relation is strictly valid only at first order, i.e. when $\Delta \mu
\ll k_{\rm B} T$.  
An example of two typical solutions of equation (\ref{eq:ell}) is
given in Fig.~\ref{fig:Dmu}, which shows $\Delta \mu / k_{\rm B} T$
for a non-superfluid star with mUrca processes at the temperature
$T=2\times 10^{8}$K, and for both model A (left panel) and B (right
panel).  For these particular cases, the `linear' approximation is
still valid for $2\times 10^{8}$K.  The temperature range, in which
this approximation holds, depends on the weak interaction process
(dUrca or mUrca), on the state of core nucleons (normal or
superfluid/superconducting) and on the magnetic field strength.  In
the following sections, we will provide the results only for models in
which $\Delta \mu \ll k_{\rm B} T$ is satisfied.

\begin{figure*}
\begin{center}
\includegraphics[trim = 42mm 12mm 38mm 10mm, clip, height=59mm]{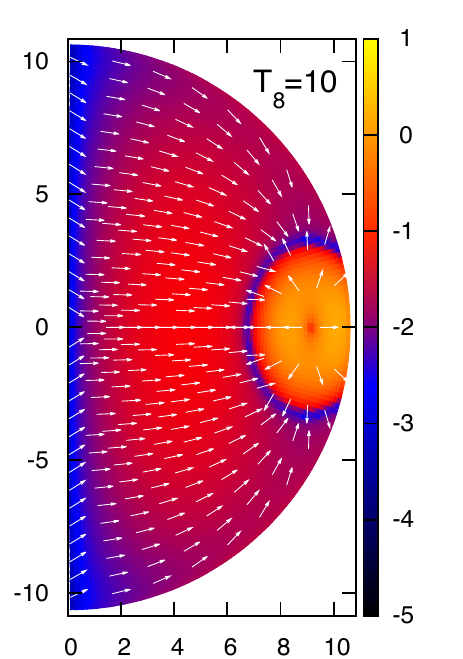} 
\includegraphics[trim = 42mm 12mm 38mm 10mm, clip, height=59mm]{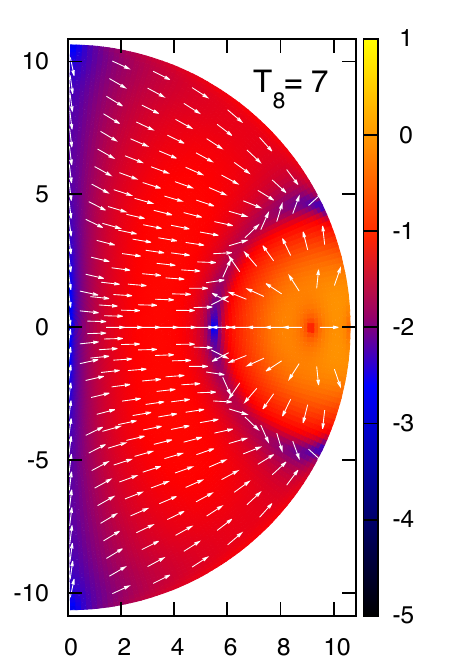}  
\includegraphics[trim = 42mm 12mm 38mm 10mm, clip, height=59mm]{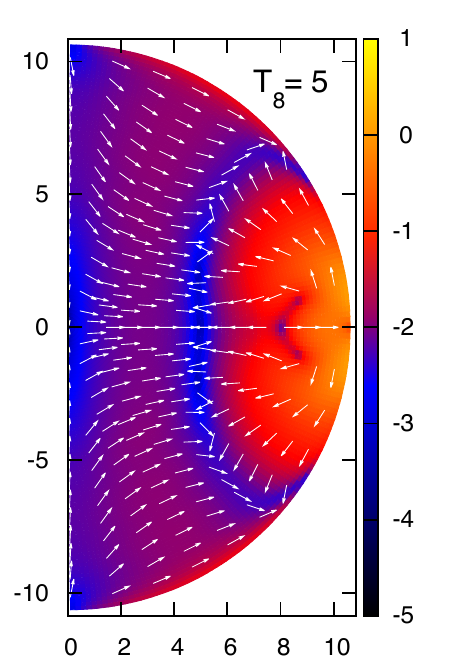} 
\includegraphics[trim = 42mm 12mm 38mm 10mm, clip, height=59mm]{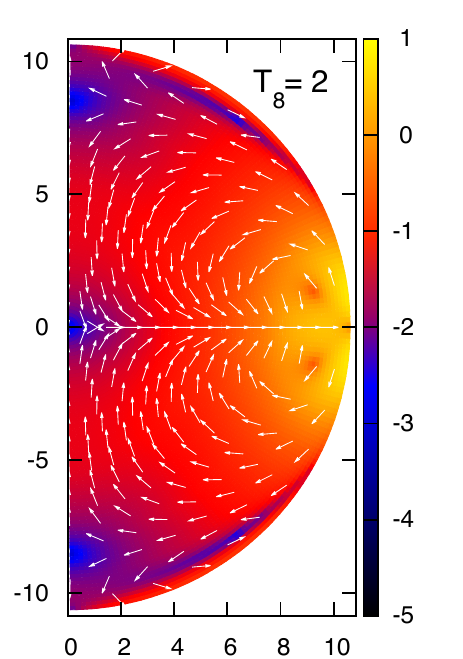}  
\caption{Same as Fig. \ref{fig:Vamb-MUrca}, for $T_{8} = 10, 7, 5$ and
  $2$, but for a magnetic field described by model B (mixed
  poloidal-toroidal magnetic field).
  \label{fig:MUrca-BpBt}}
\end{center}
\end{figure*}

\subsection{Normal matter} \label{sec:NM}

We begin our analysis with a non-superfluid/non-superconducting
neutron star core, where the weak interactions occur only through the
mUrca process. For the purely poloidal case (model A), we show in
Fig. \ref{fig:Vamb-MUrca} the 2D pattern of the ambipolar velocity for
a selection of four different temperatures, respectively, $T_8 = 10,
4, 3$ and 2 (where $T_8$ is the temperature in units of $10^8$K).  At
high temperature, the chemical reactions are very fast and $\Delta
\mu$ is negligible. The ambipolar velocity is proportional to the
Lorentz force, and it exhibits a dominant irrotational pattern with
the flow advecting the magnetic field away from the axis, and locally
converging toward the nodal line of the poloidal magnetic field.  As
the temperature decreases, when the mUrca processes are not fast
enough to establish $\beta$-equilibrium, the chemical gradients
partially cancel the Lorentz force, more precisely, the irrotational
component of the $\mvb{f}_{\! mag} / n_{\rm c}$ vector (see the
Helmholtz-Hodge decomposition in Appendix \ref{sec:app1} for more details), and
the velocity pattern is modifed.  The transition from an
irrotational-dominated flow to a solenoidal-dominated flow (see the
third and fourth panel of Fig. \ref{fig:Vamb-MUrca}) is clearly
observed, with two vorticity zones in each hemisphere, a narrower one
close to the crust/core interface and a second wider zone in the
interior.  As expected from the temperature dependence of
$\tau_{\p\n}$, the speed of the
ambipolar diffusion is larger at lower temperatures. In particular, we
find that the ambipolar flow is faster near the crust/core interface
where one of the vorticity zones is present.

The results for the mixed-magnetic field (model B) are shown in
Fig.~\ref{fig:MUrca-BpBt}.  The qualitative properties of the velocity
pattern are similar to the purely poloidal case: transition from a
(high $T$) non-solenoidal flow to a (low $T$) solenoidal flow.
However, there are also some interesting differences.  First, the
largest speed is now reached in the toroidal magnetic field region,
which simply reflects our choice $B_{t}>B_{p}$.  More interestingly,
combined with the advection of poloidal field lines away from the
axis, we see the expansion of the region containing the toroidal
field. In some regions, these two flows are opposite, which in a real
evolution model should result in a compression of magnetic field
lines. This pattern structure is particularly evident at high
temperature (see the first panel from the left of
Fig. \ref{fig:MUrca-BpBt}).  For lower $T$, the `toroidal' flow
extends further toward the stellar interior (second and third panel of
Fig. \ref{fig:MUrca-BpBt}), until the chemical gradients become strong
enough to balance the irrotational part of $\mvb{f}_{\! mag} / n_{\rm
  c}$.  At $T_{8}=2$ a clear solenoidal flow emerges again, mainly in
the region with the toroidal magnetic field.

\begin{figure}
\begin{center}
\includegraphics[trim = 42mm 12mm 38mm 10mm, clip, height=58mm]{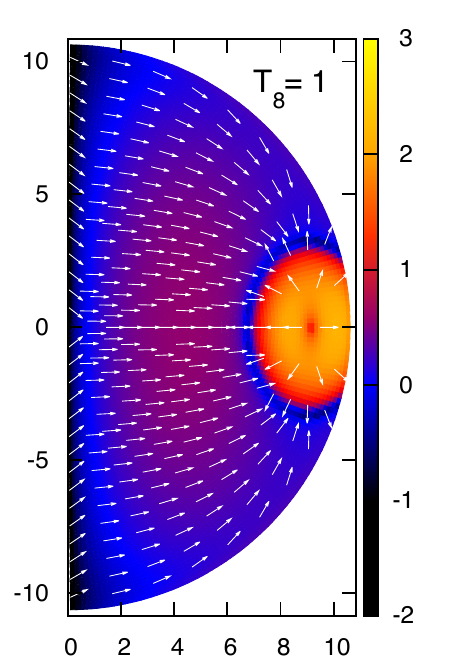} 
\includegraphics[trim = 42mm 12mm 38mm 10mm, clip, height=58mm]{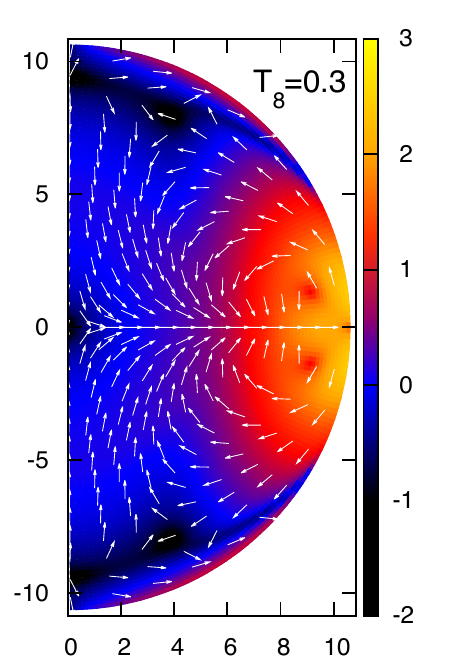} 
\caption{Same as Fig. \ref{fig:Vamb-MUrca}, for $T_{8} = 1$ and $0.3$,
  but with the magnetic field of model B and for a neutron star core
  with dUrca reactions.
\label{fig:dU}}
\end{center}
\end{figure}

If the central density of the star is sufficiently high to allow the
dUrca channel (or other fast neutrino processes), $\beta$-equilibrium
is quickly re-established. In Fig. \ref{fig:dU} we show the results
for normal matter with dUrca processes and mixed-magnetic field (model
B). This should only happen in very massive stars, and in a fraction
of the core volume, but we prefer to show results with the same
neutron star model and considering fast neutrino reactions in the
whole core to better illustrate the differences.  At $T_8=1$, the
irrotational pattern of the flow is still dominant, while at lower
temperatures $\Delta \mu$ begins to affect the solution.  The
transition to a mainly solenoidal velocity pattern occurs for
temperatures below $4 \times 10^7$K (see the right-hand panel of Fig.
\ref{fig:dU}).  Note that the flow speed is now much higher, reaching
values of about $10^2$-$10^3~\textrm{km/Myr}$ in some regions.  This
means that ambipolar diffusion can have an important effect in more
massive neutron stars, on timescales of kyr. However, in the fast
cooling scenario with dUrca processes, the star also cools much
rapidly and it remains to be proven by detailed simulations if
significant magnetic diffusion can occur before the star becomes too
cold.

\subsection{Superfluid/superconducting matter} \label{sec:SM}

A realistic neutron star is expected to become superfluid and
superconducting, resulting in very different timescales compared to
the normal matter case. This is mainly due to the suppression of the
proton and neutron collision and $\beta$-reaction rates, especially in
the strong superfluid regime. As described in Sec.~\ref{sec:sup0}, we
include the superfluid/superconducting correction on the reaction
rates and replace the Lorentz force with the superconducting magnetic
force.

To avoid the coexistence of normal and superconducting regions inside
the star and thus in our numerical domain, we consider constant gap
models, i.e. $T_{c\x}$ independent of density. In fact, it is not
clear how to handle, macroscopically, regions where the magnetic force
changes from normal to superconducting states. This transition is
likely not sharp and occurs in an intermediate layer where
superconducting fluxtubes should gradually join the magnetic field in
a normal state. Essentially, in our model, we use the Lorentz force
when $T> T_{c\p} $ and the superconducting force when $T \leq T_{c\p}
$. For the superfluid/superconducting case we discuss only the
mixed-magnetic field configuration given by model B.

\begin{figure}
\begin{center}
\includegraphics[trim = 42mm 12mm 38mm 10mm, clip, height=58mm]{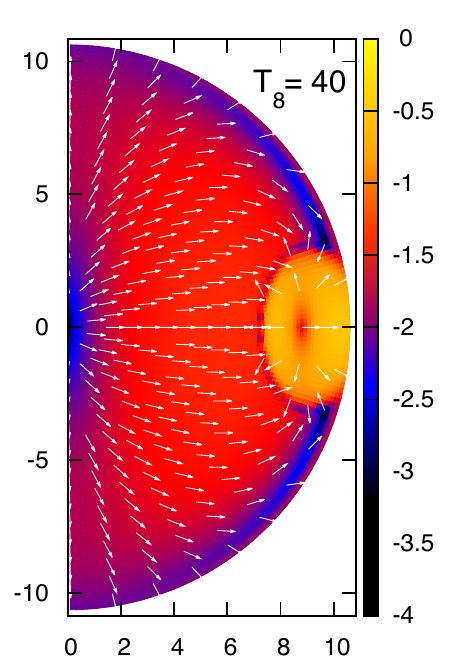} 
\includegraphics[trim = 42mm 12mm 38mm 10mm, clip, height=58mm]{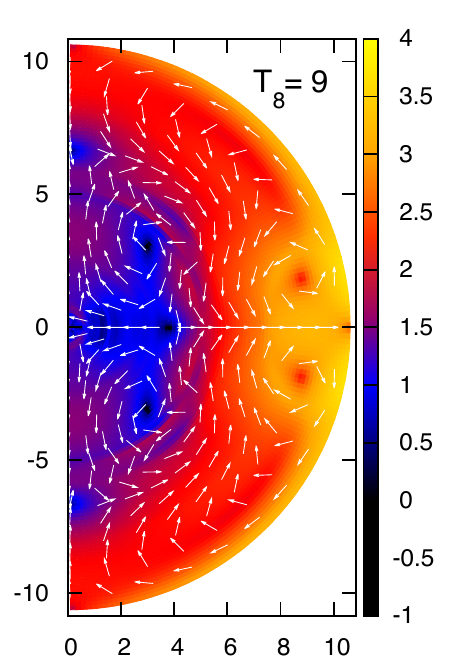} 
\caption{Same as Fig. \ref{fig:Vamb-MUrca}, for a
  superfluid/superconducting star with the magnetic field described by
  model B and mUrca processes. We show results for $T_8=40$ and 9,
  from left to right.
\label{fig:mU-Sf-Sup}}
\end{center}
\end{figure}

In Fig. \ref{fig:mU-Sf-Sup} we show the results for a neutron star
with mUrca reactions and with critical temperatures given by $T_{c \p}
= 5 \times 10^9$K and $T_{c \n} = 10^9$K, respectively.  This choice
is consistent with the theoretical calculations, which predict a
higher transition temperature for protons. During the cooling of a
neutron star, we therefore expect that the superconducting transition
sets in earlier than the superfluid transition of neutrons.  One of
the main effects of the superconducting/superfluid transition is that
the suppression of the irrotational part of the Lorentz force by the
chemical gradients occurs at higher temperature with respect to the
normal case, because of the longer reaction rate of the
$\beta$-equilibrium processes.  Comparing Fig.~\ref{fig:mU-Sf-Sup} to
Fig. \ref{fig:MUrca-BpBt}, we clearly see that when the temperature is
$T_8 = 9$ the ambipolar velocity is already dominated by the
solenoidal mode, with two large vorticity zones.  More interestingly,
the ambipolar flow now reaches a very high speed of $\sim
10^3~\textrm{km/Myr}$.  Differently from the non-superfluid case, the
maximum of $v_{amb}$ is not only restricted into the closed field line
region but it is large also outside, in a wide spherical shell. This
effect is in part due to the different form of the magnetic force,
that scales with $H_{c1} B$ instead of $B^2$.

Certainly, the temperature at which the solution becomes mostly
solenoidal depends on the particular choice of $T_{c \p}$ and $T_{c
  \n}$.  We have explored different critical temperatures $T_{c \x}$,
the ambipolar diffusion pattern is similar to what just described, but
the transition to a solenoidal velocity occurs at different $T$.  We
will return to this point in the next section.

\begin{figure}
\begin{center}
\includegraphics[trim = 42mm 12mm 38mm 10mm, clip, height=58mm]{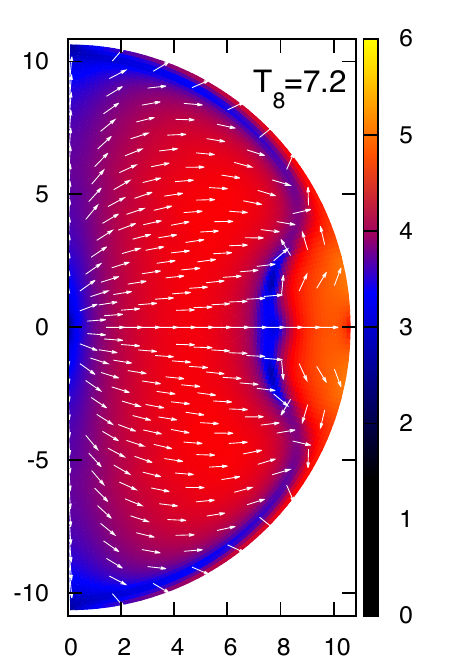} 
\includegraphics[trim = 42mm 12mm 38mm 10mm, clip, height=58mm]{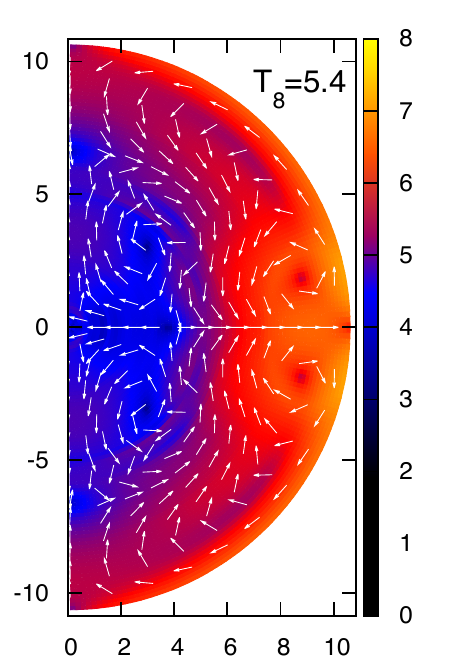} 
\caption{Same as Fig. \ref{fig:mU-Sf-Sup} but for a star with dUrca
  reactions, and $T_{8} = 7.2$ and 5.4.
\label{fig:dU-Sf-Sup}}
\end{center}
\end{figure}

If dUrca reactions are activated, the chemical gradients begin to
balance the irrotational part of $\mvb{f}_{\! mag} / n_{\rm c}$ when $
T_8 \lesssim 8$. 
We show in Fig.~\ref{fig:dU-Sf-Sup}, the velocity pattern for two
cases, respectively, at $T_8=7.2$ (left-hand panel) and $T_8=5.4$
(right-hand panel). In the former case, $\Delta \mu$ begins to affect
the velocity, while in the latter the characteristic vorticity zones
become visible.  The most interesting result is that this particular
case (superfluid/superconducting star with fast neutrino cooling
processes) results in the largest velocities, up to $10^6-10^8$~km/Myr
in the temperature interval $ 5 < T_8 < 7$. The impact on the magnetic
field evolution is therefore potentially strong. The rapid cooling
induced by dUrca reactions can however moderate the effects of this
high speed. A conclusive answer can be only given when the ambipolar
drift is consistently incorporated in simulations of the
magneto-thermal evolution. This issue will be addressed in a future
work.
\section{Timescales} \label{sec:timescale}

We now discuss the timescales associated with ambipolar diffusion and
compare our results with the analytical estimates given in the
literature. From the numerical solutions we determine two different
timescales. The first is defined by the following equation:
\begin{equation}
t_{amb} = \frac{L}{\langle v_{amb}\rangle} \, , \label{eq:tev}
\end{equation}
where $L$ is a typical distance in which the magnetic field varies,
and $\langle v_{amb}\rangle$ is the volume average of the velocity
modulus. Equation (\ref{eq:tev}) provides an average timescale on
which a magnetic field line is advected to a distance $L$ by a
velocity $v_{amb}$. This may or may not result in field dissipation.
To study the magnetic field dissipation rate, we also introduce the
following timescale:
\begin{equation}
t_{B} = - \frac{2 E_{B}}{ \dot E_{B} } \, , \label{eq:tdis}
\end{equation}
where $E_{B}$ and $\dot E_{B}$ are, respectively, the magnetic energy
and the energy dissipation rate due to ambipolar velocity (the dot
denotes a time derivative).  For normal matter, these quantities read
\citep{1992ApJ...395..250G}
\begin{align}
   E_{B} & = \frac{1}{8\pi} \int dV B^2 \, , \\ \dot E_{B} & = - \int
   dV \mvb{v}_{amb} \cdot \mvb{f}_{\! mag} \, .
 \label{edot}
\end{align}
To determine equation (\ref{eq:tdis}), we have assumed that the time
dependence of the magnetic field is $B \sim e^{-t/t_{B}}$. We note
that a large $v_{amb}$ that is nearly perpendicular to the Lorentz
force gives a fast evolution of the magnetic field, $t_{amb}$, but
without dissipation.  For superconducting stars we do not determine
the timescale $t_B$, as the analogous to Eq. (\ref{edot}) contains
also surface terms which are not negligible. These terms strongly
depend on the matching conditions imposed at the interface separating
the superconducting and normal states.

\begin{figure*}
\begin{center}
  \includegraphics[height=80mm]{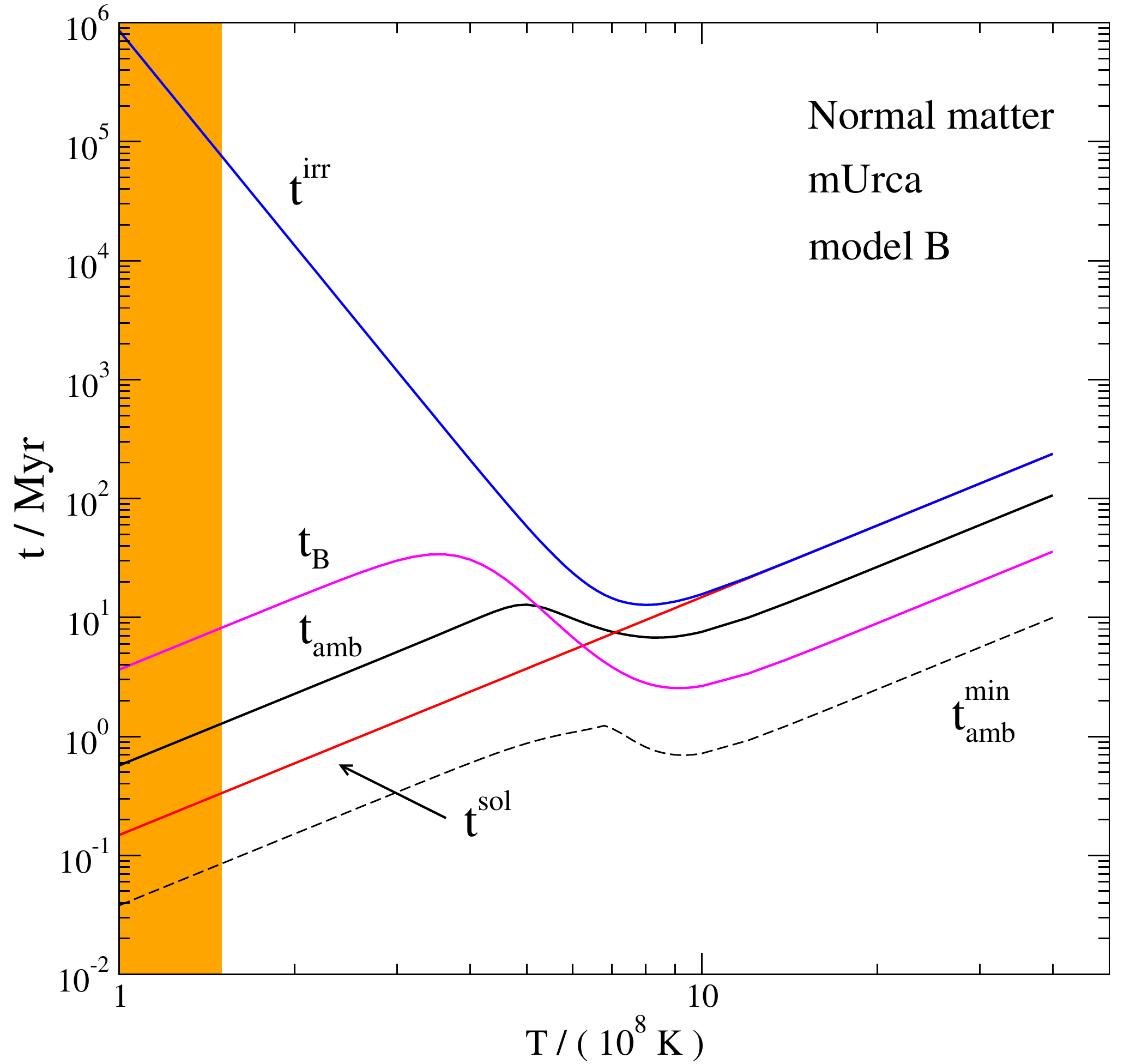} 
  \includegraphics[height=80mm]{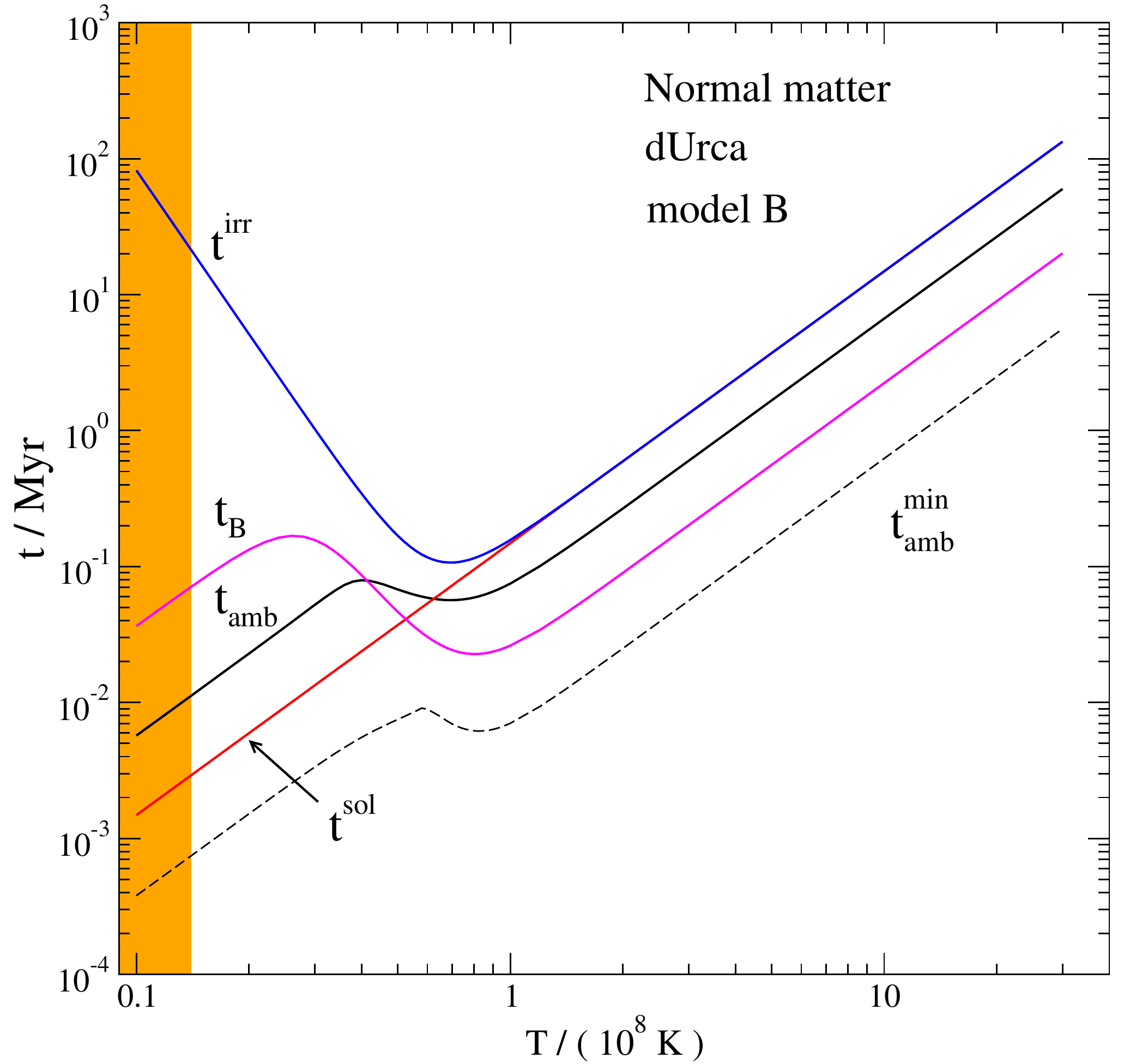} 
\caption{Ambipolar diffusion timescales as a function of temperature
  for normal matter and model B. The left panel shows results assuming
  mUrca reactions and the right panel for dUrca processes. The black,
  red and blue solid lines, respectively, denote the volume average of
  the $t_{amb}$, $t^{sol}$ and $t^{irr}$ timescale for $L=1$~km. The
  black-dashed line refers to the minimum of $t_{amb}$, while the
  magenta-solid line shows $t_B$. The orange region delimits the
  temperature range where the condition $\Delta \mu \ll k_B T$ is not
  satisfied.
\label{fig:t-NM-mUrca-bt0}}
\end{center}
\end{figure*}

In our numerical approach, we do not separate explicitly between
solenoidal and irrotational components, therefore $t_{amb}$ and
$t_{B}$ describe global timescales.  By using the Helmholtz-Hodge
decomposition of the velocity and magnetic force,
\citet{1992ApJ...395..250G} found the following analytical estimates
for the solenoidal and irrotational ambipolar diffusion timescales:
\begin{align}
&  t^{sol} \sim \frac{4\pi m^{*}_{\p} n_{\rm c} L^2}{\tau_{\p \n} B^2 }  \, , \qquad 
t^{irr} \sim t^{sol}  \left( 1 + \frac{a^2}{L^2} \right) \, ,  \label{eq:tscale}
\end{align}
where $a$ is the coefficient defined in equation
(\ref{eq:ab})\footnote{In \citet{1992ApJ...395..250G}, the coefficient
  $a$ is defined as in Eq. (\ref{eq:ab}) but with $x_{\n}=1$.}, and we have replaced   
   $m_{\p}$ with a effective proton mass $m_{\p}^{*}$.     At
high temperature, $a \ll L$ and the two timescales are almost the
same.  At low $T$, the chemical gradients suppress the irrotational
part of the force and $t^{sol} \ll t^{irr}$. To calculate these
analytical quantities, we specify a typical $L$ and determine
$t^{irr}$ and $t^{sol}$ in the entire numerical grid, and then extract
either their minimum or their volume average value.

In Fig.~\ref{fig:t-NM-mUrca-bt0}, we show the temperature dependence
of the ambipolar diffusion timescales for a stellar model with normal
matter, mUrca reactions, and a magnetic field described by model B.
We use $L=1$~km to determine the volume average of $t_{amb}$,
$t^{sol}$ and $t^{irr}$. The two analytical timescales almost
coincide, as expected, at high temperature, while they start to
diverge when $T_8 \lesssim 10$. The solenoidal timescale reflects the
$T^2$ dependence of $\tau_{pn}^{-1}$, while the irrotational
timescale, when $a \gg L$, becomes independent of $L$ and scales as
$\lambda^{-1}$ ($T^{-6}$ and $T^{-4}$ for mUrca and dUrca processes,
respectively).

Our numerical results agree, within an order of magnitude, with the
analytical estimates, but with some interesting differences. The
numerical $t_{amb}$ shows the correct temperature scaling ($T^2$) at
high and low temperature, with a bump indicating the transition to
solenoidal flow at $5 < T_8 < 10$.  Note that the large timescales
predicted by the irrotational mode estimates at low temperature will
never be realized in a real scenario. That would require to construct
a magnetic field configuration such that the Lorentz force per charged
particle is a purely irrotational vector. In any other case, there
will be a non-vanishing solenoidal part that determines the actual
timescale of ambipolar diffusion.  We only see the temperature scaling
corresponding to the irrotational mode in the transition temperature
interval, while $\Delta \mu$ is growing to balance the irrotational
part of the $\mvb{f}_{\! mag} / n_{\rm c}$ vector.  However, the
presence of a solenoidal part limits the increase of $t_{amb}$ and,
for $T_8 < 5$, it follows again the $T^2$ scaling.  The transition to
a predominant solenoidal solution always shows this characteristic
S-shape.  The dissipation timescale $t_B$ is also shown in
Fig.~\ref{fig:t-NM-mUrca-bt0}. It follows the same qualitative
behaviour as $t_{amb}$ but with a wider variation. At high temperature
the velocity and the magnetic force are always parallel, which
maximizes $\dot E_{B}$, resulting in short diffusion timescales.  At
low $T$, when the solenoidal flow dominates, the Lorentz force and the
velocity field are no longer aligned, which explains why at $T_8<3$ we
find $t_{B} > t_{amb}$.

For the same stellar model we also consider the case of dUrca
reactions (right panel). The results are similar, except that the
transition now appears at $0.4 <T_8 < 0.8$, at a temperature which is
roughly one order of magnitude smaller than the mUrca case, due to the
enhanced efficiency of the $\beta$-reaction rates.

Our results show that for the mUrca case, in the temperature range $1
\leq T_8 \leq 10$, $t_{amb}$ is larger than 1~Myr.  This value
decreases by an order of magnitude if we consider the minimum
numerical timescale $t_{amb}^{min}$ (shown as a dashed-line in
Fig.~\ref{fig:t-NM-mUrca-bt0}), but this is only the minimum value
reached locally in the star core, with little relevance to the overall
evolution. We can safely conclude that ambipolar diffusion is
irrelevant during the first Myr of a neutron star life, for normal
matter, in the standard cooling scenario, and magnetic fields $B \le
10^{14}$~G. If dUrca processes are activated, the timescales are
reduced considerably. The minimum timescale of our numerical solutions
can reach $t_{amb}^{min} \simeq 1$~kyr while the global quantities
$t_{amb}$ and $t_B$ reach values as low as 10 kyr, comparable to the
expected ages of young X-ray pulsars.  The imprint of fast neutrino
cooling processes could be, in principle, visible as a fast magnetic
field evolution in the core, driven by ambipolar diffusion. However,
we need to incorporate superfluid/superconducting effects to be closer
to the real case, which is done in the next section.

\begin{figure}
\begin{center}
\includegraphics[height=80mm]{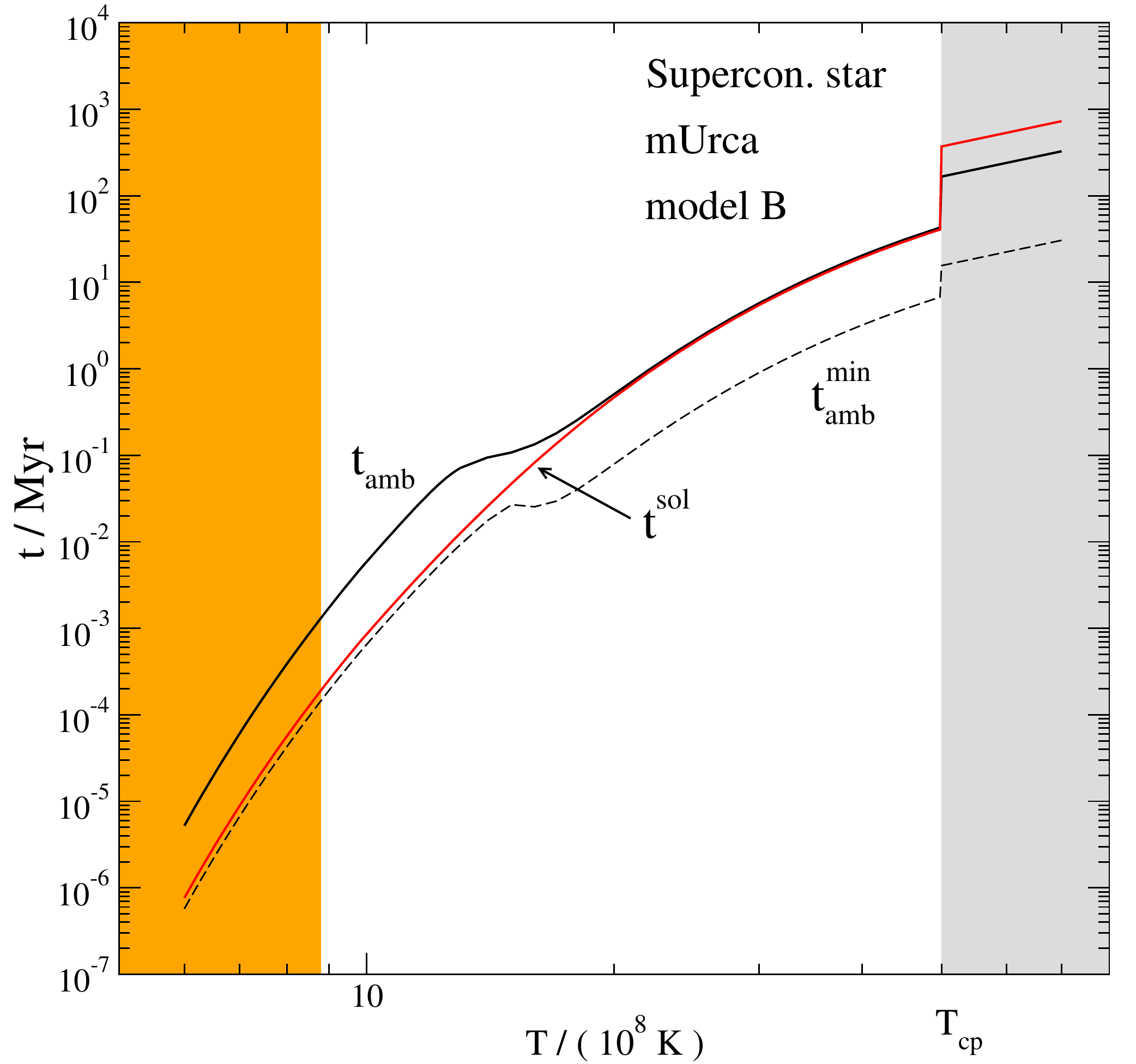}
\caption{Variation of the ambipolar diffusion timescale with the
  temperature.  Differently from the case shown in
  Fig.\ref{fig:t-NM-mUrca-bt0}, the star is superconducting. 
  The $\beta$-reactions are driven by the mUrca process.  The proton
  critical temperaturer are $T_{c\p}=5\times 10^9$~K and $T_{c\n}=
  10^9$~K. The black-solid line denotes the $t_{amb}$ timescale, while
  the back-dashed line its minimum $t_{amb}^{min}$. The red-solid line
  refers to the volume average of the analytical solenoidal timescale
  $t^{sol}$ (Eq.~\ref{eq:ts-SF}). The grey region denotes the temperature in which the 
  star is not superconducting. 
\label{fig:t-SF-mUrca-bt15}}
\end{center}
\end{figure}

\subsection{Superconductivity} \label{sec:sup-time}

When the star is superconducting or superfluid the ambipolar diffusion
timescales may be very different than the normal matter
case. \citet{2011MNRAS.413.2021G} derived analytical estimates for the
solenoidal and non-solenoidal motion, which are given by
\begin{align}
& t^{sol} \sim \frac{4\pi m^{*}_{\p} n_{\rm c} L^2}{H_{c1} B }
  \frac{\mathcal{R}_{\n\p}}{\tau_{\p \n}} \, , \qquad t^{nsol} \sim
  \frac{4\pi n_{\rm c}^2 }{\lambda \mathcal{R}_{sf} H_{c1} B } \,
  ,  \label{eq:ts-SF}
\end{align}
 where we have replaced $m_{\p}$ with an effective proton mass $m_{\p}^{*}$.   
These timescales are determined by assuming that the particle
scattering is dominant over the mutual friction dissipation
(interaction between vortices and fluxtubes), which is approximately
correct when $T_8 \gtrsim 3$.

In Fig. \ref{fig:t-SF-mUrca-bt15}, we show the results for a
superconducting/superfluid neutron star with mUrca reactions and a
mixed magnetic field (model B). The proton critical temperature is
$T_{c\p} = 5\times 10^9$~K, while the neutron transition to
superfluidity is at $T_{c\n}=10^9$~K. After the superconducting
transition the numerical timescale $t_{amb}$ closely follows the
analytical estimate $t^{sol}$. The bump associated with the transition
to solenoidal flow is now smaller than in the non-superconducting
case. The effect of the $\mathcal{R}_{\n\p}$ correction factor on the
collision times results in a temperature dependence somewhat different
from the $T^2$ scaling observed in the normal matter case.

\begin{figure*}
\begin{center}
\includegraphics[height=80mm]{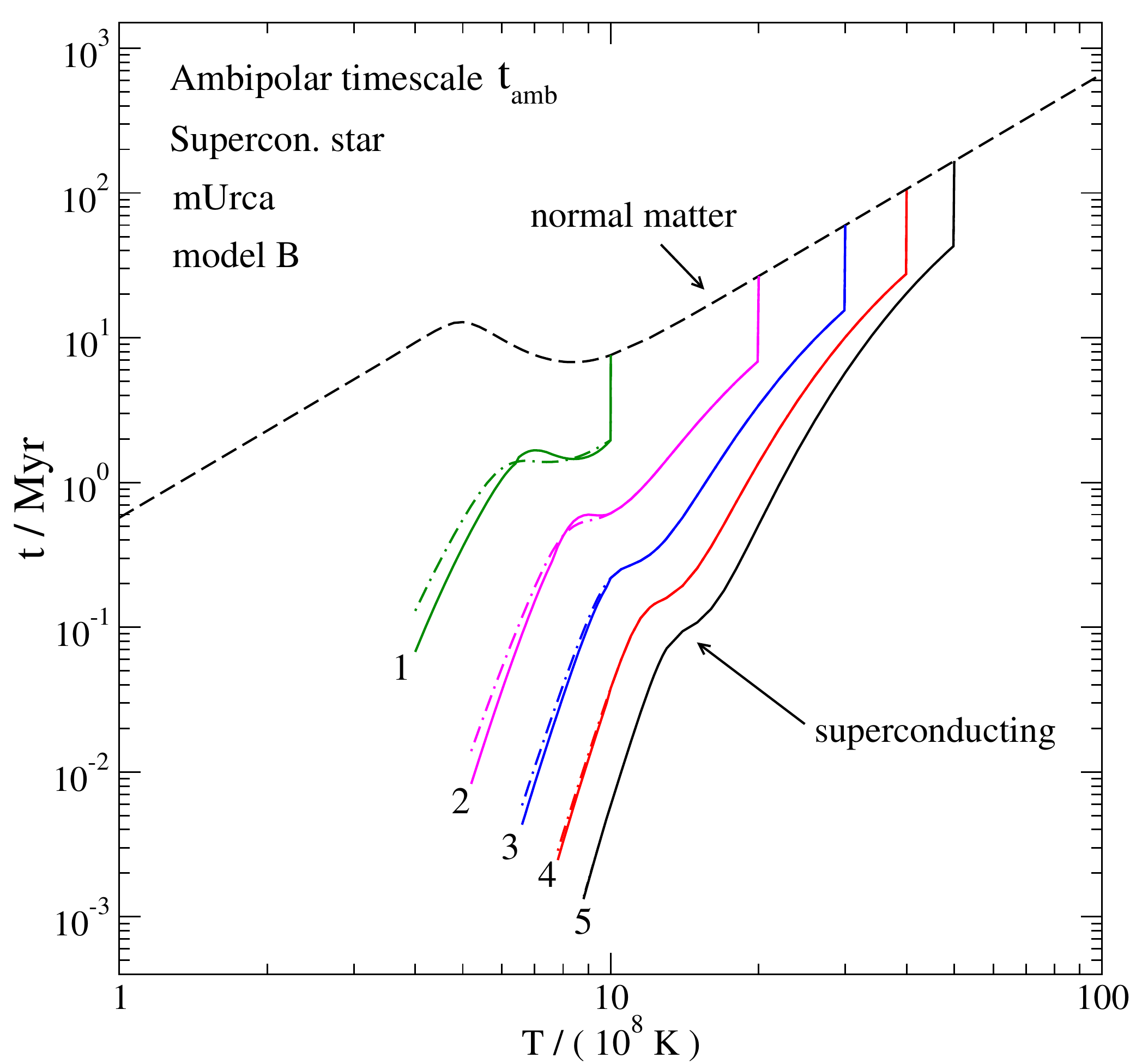}
\includegraphics[height=80mm]{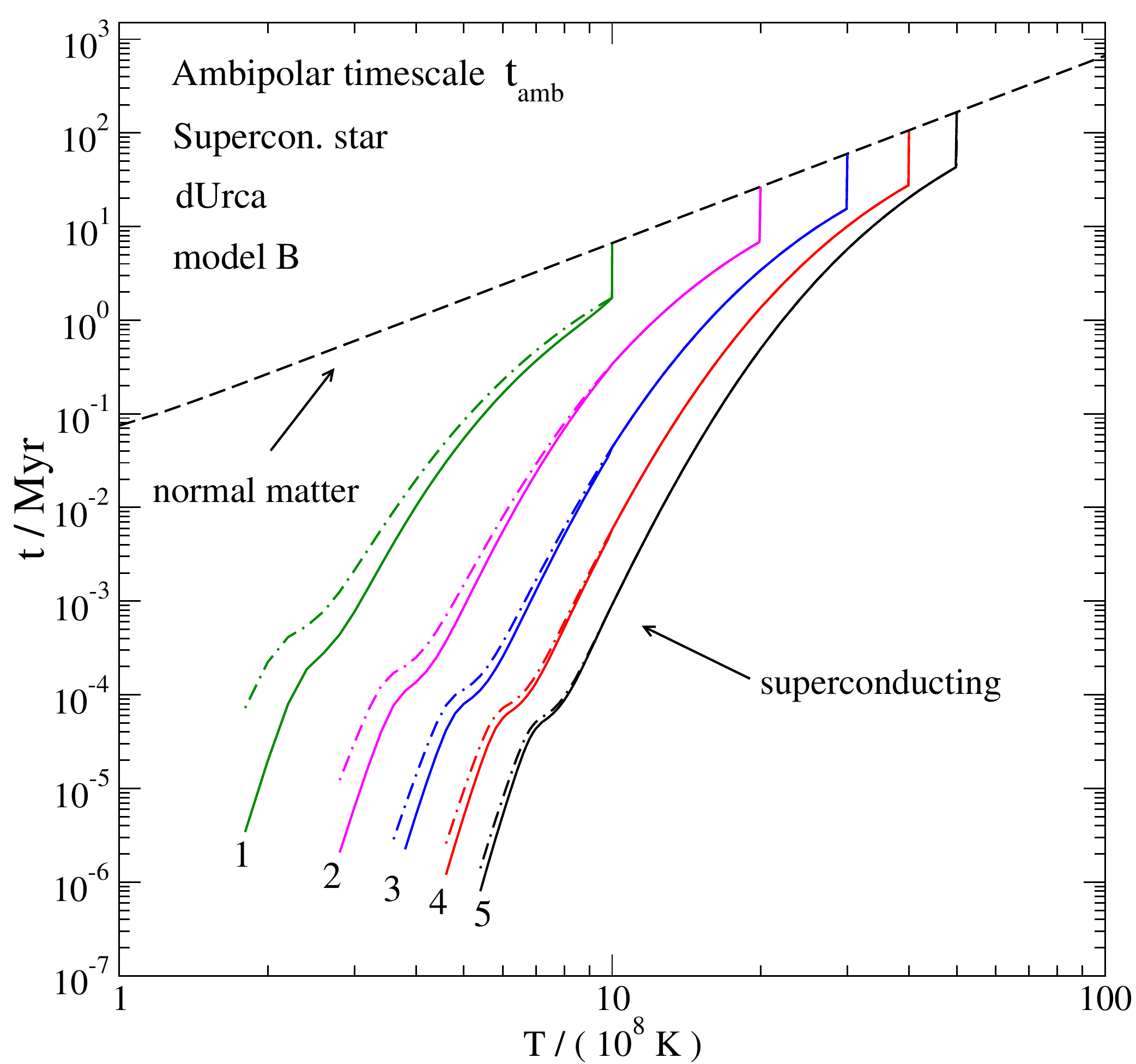}
\caption{Ambipolar diffusion timescale $t_{amb}$ as a function of
  temperature for superconducting models with different critical
  temperature $T_{c\p}$, and with mixed-magnetic field (model B). The
  left- and right-hand panels show, respectively, the results for
  stars with mUrca and dUrca reactions.  The dashed-line denotes the
  $t_{amb}$ of a normal matter star, the colored-solid lines describes
  the $t_{amb}$ for superconducting stars with $T_{c\n}=10^9$~K, while
  the dot-dashed lines refer to a model with normal neutrons.  The
  number at the end of the superconducting $t_{amb}$ denotes to the
  proton critical temperature, namely $T_{c\p}=1,2,3,4,5 \times
  10^9$~K.
\label{fig:t-SF-Tcp}}
\end{center}
\end{figure*}

Finally, we explore the effects of different proton critical
temperatures in both mUrca and dUrca scenarios. The results are
summarized in Fig. \ref{fig:t-SF-Tcp} , where we show the timescales
for $T_{c\p} =1, 2, 3, 4, 5 \times 10^9$~K for $T_{c\n}=10^{9}$~K
(solid lines) and without neutron superfluidity (dot-dashed
lines). The results show the little effect of the superfluid neutron
transition, provided that the protons become superconducting at higher
temperature.  Minor differences are only visible when the two critical
temperatures are similar $T_{c\n} \simeq T_{c\p}$.  As expected, the
temperature at which the transition to a solenoidal flow occurs,
depends on the critical temperature $T_{c\p}$ and gradually increases
for higher $T_{c\p}$.
The main result in the superconducting case is that, due to the weaker
particle interactions, the global evolution timescales are sensibly
reduced, being as short as 1-10 kyr for the mUrca case, or even of the
order of years for the dUrca case.  Our results show that when the
critical temperature $T_{c\p}$ is higher, ambipolar diffusion can have
a more significant impact on the magnetic field evolution. We must
note again that in the dUrca case, cooling of the star also proceeds
much faster, and a more detailed study is needed before reaching more
robust conclusions. But there is a potentially large effect of the
proton superconducting gap on the core magnetic field evolution, which
can be used to constrain its value through the combination of detailed
modelling and astrophysical observations.

\section{Conclusions} \label{sec:con}

We have revisited the problem of ambipolar diffusion in neutron stars
with axisymmetric magnetic fields, with special attention to the
relevance of microphysical details (fast versus slow neutrino
processes, normal versus superfluid matter).  For a given magnetic
field configuration and temperature, we determine numerically the
local deviations from $\beta$-equilibrium and the relative velocity of
the charged component (protons and electrons) with respect to the
neutral component (neutrons), which causes the diffusion of the
magnetic field.

In the wide range of temperatures and parameters explored, we could
follow the variation of the velocity field and identify the
temperature interval in which a solenoidal pattern becomes dominant in
the flow. This transition to a solenoidal solution is due to the
effect of the small departure from chemical equilibrium, which results
in local pressure gradients that balance the irrotational part of the
``magnetic force'' acting on charged particles.  The temperature at
which this transition occurs depends on the $\beta$-reaction rates,
superfluid/superconducting gap models, etc.
 
Typical core temperatures of neutron stars are between $10^8$~K and
$10^9$~K, depending on the age and efficiency of neutrino
reactions. Ambipolar diffusion can influence the evolution of the core
magnetic field if, in this temperature interval, its timescale is of
the order of the star age (from $10^3$ to $10^6$ yr). We find that
such relatively short timescales can be achieved at low temperatures,
after the transition to a solenoidal flow. However, in all cases
dominated by an irrotational flow, ambipolar diffusion is expected to
have little effect, as the magnetic field evolves on longer
timescales.
 
For stars composed of normal matter with $\beta$-reactions controlled
by the mUrca processes, the shortest evolution timescale is about
1~Myr at $T\approx 10^8$~K, for a mixed magnetic field configuration
with $B_p = 10^{14}$G at the magnetic pole and a maximum toroidal
field of $10^{15}$~G.  If the dUrca process is activated, shorter
ambipolar diffusion timescales of the order of $10$ kyr are reached at
$T=1-3 \times 10^7$~K. This could be the case of more massive NSs,
where the central density is higher and additional neutrino channels
could be opened.

However, NS cores are expected to be superfluid and superconducting
and the suppression of both particle collisions and weak interaction
rates substantially changes the results. In superfluid/superconducting
cores, we find that, at about $T\approx 10^9$~K, the ambipolar drift
timescale is about 1 kyr for mUrca processes and can be as short as
about few years in stars with dUrca processes. The temperature at
which the ambipolar flow reaches these timescales depends on the
critical temperatures of superconducting and superfluid transitions.
Our results show that ambipolar diffusion can play a key role in the
magnetic field evolution in the superconducting core of a neutron
star.  However, there are uncertain aspects of the physical processes
in this conditions that need to be carefully revised, in particular
the interactions between particles and fluxoids. The most interesting
cases are when dUrca reactions are present, but in this situation the
star also cools much faster, so it is unclear whether a substantial
modification of the magnetic field configuration has observable
consequences.

We need to go beyond the present approach that only gives information
about snapshots of the NS life, at a fixed temperature and magnetic
field configuration, and incorporate ambipolar diffusion consistently
in magneto-thermal simulations. It is also possible that the
non-linear evolution of the magnetic field brings the system quickly
into a nearly force-free configuration, that reduces the impact of
ambipolar diffusion. To firmly establish the role of the ambipolar
diffusion in the evolution of neutron stars we must rely on
multidimensional numerical simulations.  The code developed in this
work for calculating the global velocity field serves to set up the
stage for this next step.


\section*{Acknowledgements}
We would like to thank Andreas Reisenegger for his useful comments.   
A.P. acknowledges support from the European Union under the Marie
Sklodowska Curie Actions Individual Fellowship, grant agreement
n$^{\rm o}$ 656370.  This work is supported in part by the Spanish
MINECO grants AYA2013-42184-P and AYA2015-66899-C2-2-P, and by the New
Compstar COST action MP1304. 

\nocite*
\bibliographystyle{mn2e}

\appendix
\section{The equation for $\Delta \mu$} \label{sec:app0}

In this section, we present the derivation of equation (\ref{eq:ell}) and describe the approximation we have used. 
Taking the divergence of Eq. (\ref{eq:vamb}) and using the continuity Eqs. (\ref{eq:con1}),  we obtain
\begin{equation}
 \nabla^2  \left( \Delta \mu \right) 
-  \frac{1 }{b}   \,    \frac{ \partial \Delta \mu }{\partial r} 
-  \frac{x_n}{a^2  \lambda}  \, \m{\nabla} \cdot  \left( n_{\bb} \mvb{v}_{\n} \right)
=  \m{\nabla} \cdot  \left( \frac{ \mvb{f}_{\! mag} }{ n_{\crt}}      \right)
- \frac{1 }{b}  \, \frac{ f^{r}_{ mag} }{ n_{\crt}}    
\, ,    \label{eq:ell0App}
\end{equation}
where  we have assumed that the microphysical coefficients only depend on the radial
coordinate, and we have defined 
\begin{align}
& \frac{1}{a^2} = \frac{\lambda \, m_{\p}^* }{x_{\n}^2 n_{\crt} \tau_{\p \n} } \, , \qquad \qquad  
 \frac{1}{b} =  \frac{d}{d r} \ln \left( \frac{m_{\p}^*}{x_{\n} n_{\crt} \tau_{\p \n}}  \right) \, .
\label{eq:abApp}
\end{align}
Both $a$ and $b$ have dimensions of length.  

To determine  an equation for only  $\Delta \mu$, we must remove the neutron velocity, which requires  some further approximations. 
For instance, \cite{1992ApJ...395..250G}   neglected the proton contribution to the total
mass (i.e. $n_{\n}=n_{\bb}$, $x_n=1$), which leads  to
\begin{equation}
\m{\nabla} \cdot  \left( n_{\bb} \mvb{v}_{\n} \right) \approx \m{\nabla} \cdot  \left( n_\n \mvb{v}_{\n} \right) = \lambda \Delta \mu \, ,
\end{equation}
and thus
\begin{equation}
 \nabla^2  \left( \Delta \mu \right) 
-  \frac{1 }{b}   \,    \frac{ \partial \Delta \mu }{\partial r} 
-  \frac{1}{a^2}  \, \Delta \mu
=  \m{\nabla} \cdot  \left( \frac{ \mvb{f}_{\! mag} }{ n_{\crt}}      \right)
- \frac{1}{b}  \, \frac{ f^{r}_{ mag} }{ n_{\crt}}    
\, ,    \label{eq:ell1}
\end{equation}
where $a$ and $b$ are given by equation  (\ref{eq:abApp})  with $x_\n = 1$.

In a slightly more rigorous way, we can write 
\begin{align}
 x_{\n} \m{\nabla} \cdot  \left( n_{\bb} \mvb{v}_{\n} \right) & = x_{\n} \m{\nabla} \cdot  \left( n_{\n} \mvb{v}_{\n} / x_{\n} \right)  \nn \\
& = \lambda \Delta \mu  - n_{\bb}  \mvb{v}_{\n}  \cdot \nabla(x_{\n}) \, ,  \label{eq:appr1}
\end{align}
and assume that the last term can be neglected. This happens when $x_\n$ is constant throughout the star or when we are 
in the neutron reference frame where $\mvb{v}_{\n}=0$. With this approximation, we obtain again Eq.  (\ref{eq:ell1}), 
but with the quantities $a$ and $b$ defined by equation (\ref{eq:abApp})  with $x_\n \neq 1$.

Alternatively, we  can  write equation (\ref{eq:appr1}) as follows 
\begin{equation}
x_\n \m{\nabla} \cdot  \left( n_\bb \mvb{v}_{\n} \right) = \lambda \Delta \mu 
- (n_\bb  \mvb{v}_{\bb} - n_\p \mvb{w}_{\p\n}  ) \cdot \nabla(x_\n) \, , \label{eq:appr2}
\end{equation}
and  work in the coordinate system locally comoving with the baryons, where $n_{\bb} \mvb{v}_{\bb} = n_{\n} \mvb{v}_{\n} + n_{\p}
\mvb{v}_{\p}=0$. Neglecting $\mvb{v}_{\bb}$ in equation  (\ref{eq:appr2}), we can determine again  
equation Eq.  (\ref{eq:ell1}), where the coefficient $a$ is given by equation (\ref{eq:abApp}) while the coefficient $b$ has an extra factor $x_\n$, i.e. it now reads  
\begin{equation}
\frac{1}{b}  \to   \frac{d}{d r} \ln \left( \frac{m_{p}^*}{x_{\n}^2 n_{\crt} \tau_{\p \n}}   \right) \, .
\end{equation}

In any case, the difference between these various approximations (necessary to completely remove velocity terms from the equation) are always 
of the order of $1-x_n \approx 0.1$.

\section{Tests} \label{sec:app1}

In this section we derive an analytical solution of equation (\ref{eq:ell})
in order to test our numerical code and to understand the main
properties of the solutions.  The $\m{\nabla} \Delta \mu$ term becomes important
at low temperature ($\approx 10^8$K), when the weak interactions are
sufficiently slow. This limit corresponds to the $L/a \ll1$ case,
which leads to the following simplified version of equation (\ref{eq:ell}):
\begin{equation}
 \nabla^2  \Psi
+  \frac{1 }{b}   \,    \partial_{r}  \Psi
=  \m{\nabla} \cdot   \mvb{F}  
+ \frac{F^{r}}{b}  
\, ,    \label{eq:t1}
\end{equation}
where $\Psi$ is a function of $r$ and $\theta$, $\mvb{F}$ a general
vector field, and $b$ is a coefficient that in this section we
consider constant.  Analytical solutions can be found for specific
forms of the vector field $\mvb{F}$.

\begin{figure*}
\begin{center}
\includegraphics[height=75mm]{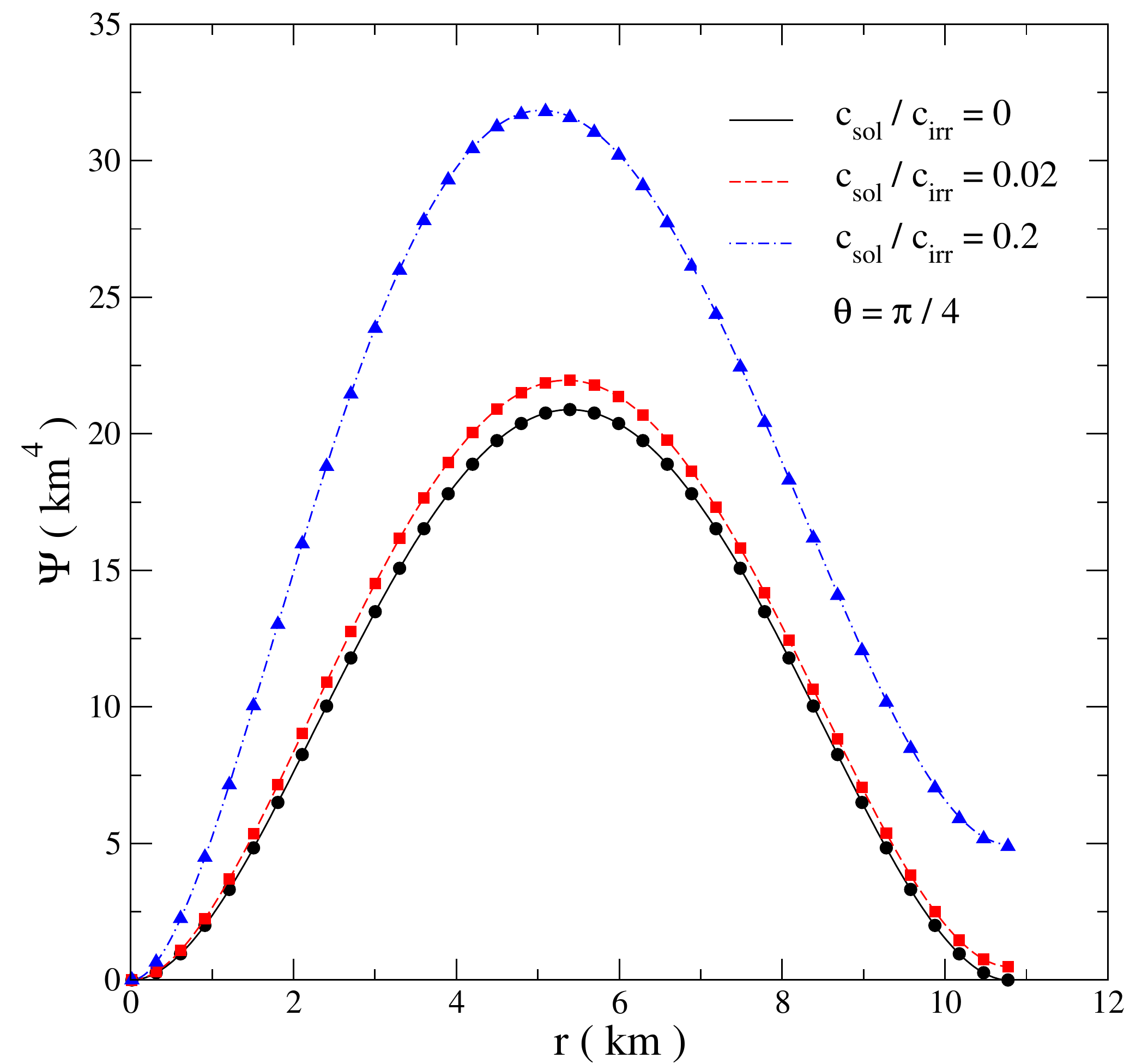}
\includegraphics[height=75mm]{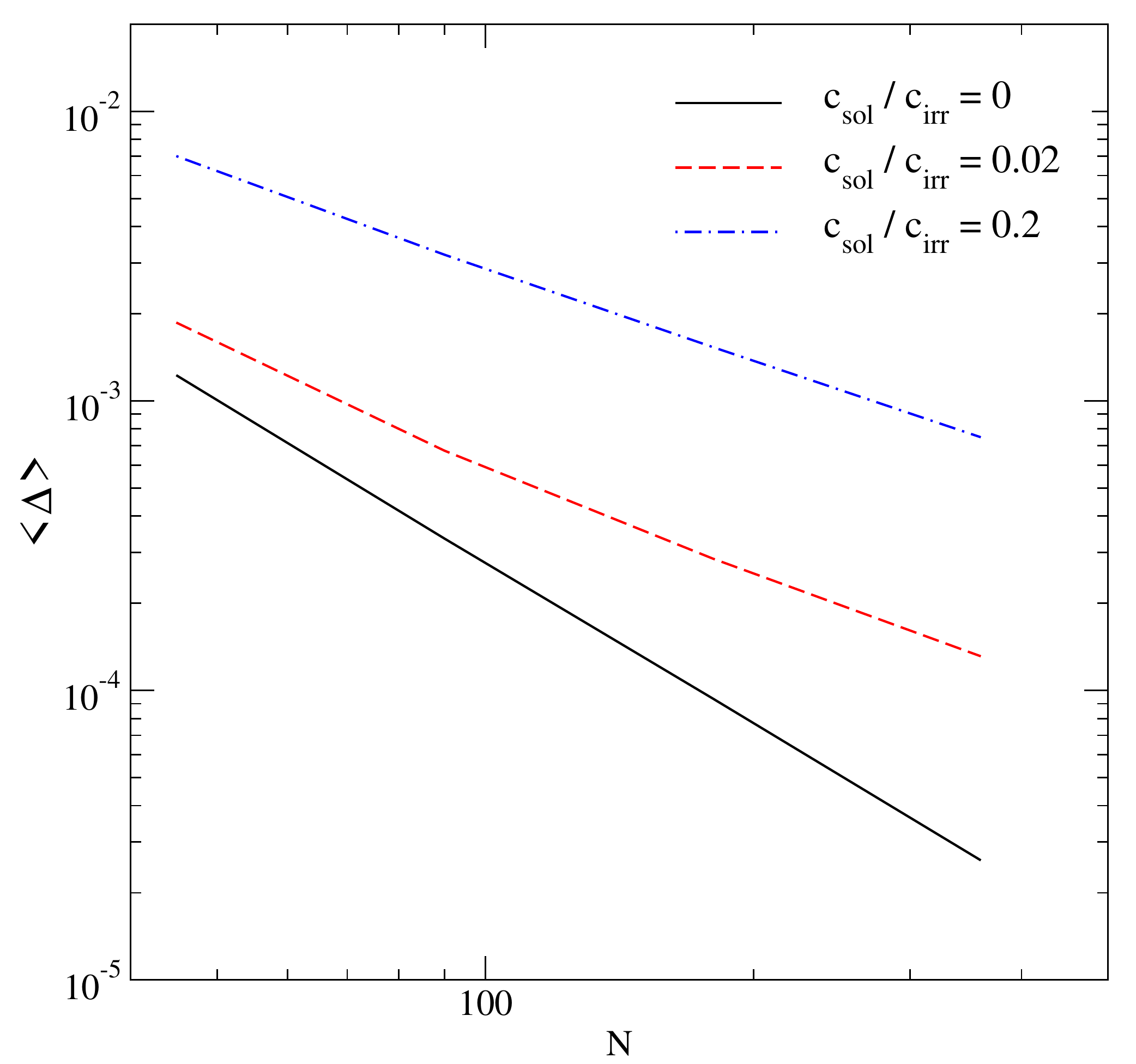}
\caption{Left-hand panel: radial profile (at $\theta=\pi/4$) of the
  numerical and analytical solution of equation (\ref{eq:t1}) for three
  cases with $c_{sol}=0$ (black-solid line), $c_{sol}/c_{irr}= 0.02$
  (red-dashed line), and $c_{sol}/c_{irr}= 0.2$ (blue-dashed-dotted
  line). The numerical solution is represented by lines, while the
  analytical solution is denoted with symbols. Right-hand panel:
  log-log plot of the averaged error $\langle \Delta \rangle$
  for various grid resolutions.  In the horizontal axis, N is the
  number of points of a 2D grid with size NxN.
\label{fig:test}}
\end{center}
\end{figure*}

In a bounded domain, with appropriate boundary conditions, we can use
the Helmholtz-Hodge decomposition to decompose $\mvb{F}$ in its
irrotational and solenoidal parts as follows
\begin{equation}
\mvb{F} = \m{\nabla} \Phi +  \m{\nabla} \times \mvb{A} \, , 
\end{equation}
where $\Phi(r,\theta)$ is a scalar function and $\mvb{A}(r,\theta)$ is
a vector field.  We consider the following expressions for the
irrotational and solenoidal parts:
\begin{align}
& \Phi = r^2 \left( r - R_a \right)^2   P_2   \, , \\
& \mvb{A} =  r^2 \left( r  - R_a \right)^2   \frac{ d P_2}{d \theta}   \m{\hat{\phi}} \, ,
\end{align}
where $R_a$ is a constant and $P_2 = \left( 3 \cos \theta ^2 - 1
\right) / 2$ is the $l=2$ Legendre polynomial.

With this choice for $\mvb{F}$, equation (\ref{eq:t1}) can be
decomposed in spherical harmonics and becomes an ordinary differential
equation in the radial coordinate.  For a vector field $\mvb{F} =
c_{irr} \m{\nabla} \Phi + c_{sol} \mvb{F}_{sol} $, where $c_{irr}$ and
$c_{sol}$ are two constants, we can derive an analytical solution
$\Psi_{an} = \psi_{an}(r) P_2(\theta) $.  The radial part of the
solution, $\psi_{an}(r)$, which is regular at the origin ($r=0$) is
given by
\begin{align}
\psi_{an}(r) = & \left[  \left(4+ \frac{r}{b} \right) \, 6 \, e^{- r/b} - 24 + \frac{18 r}{b} 
- \frac{6 r^2}{b^2} +  \frac{r^3}{b^3}  \right]  \frac{c_0}{ r^3 } \nn \\
& - \frac{3}{2}   \left[  2 R_a ^2 - \frac{8}{3}  R_a r + r^2 -  \frac{14}{3} 
 \left( -  \frac{12}{7}  R_a + r  \right)  b  \right. \nn \\
& \left.  + 14  b^2  \right] r^2 c_{sol}  
+ r ^2  \left( r - R_ a \right)^2  c_{irr}  \, .
\end{align}
The constant $c_0$ can be determined by imposing the external boundary
condition at $r=R_{a}$.  As described in Sec.~\ref{sec:res}, we
consider the outer boundary condition $ \partial_r \Psi = F^r(r=R_a)=0
$, which leads to the following expression:
\begin{align}
c_0 = \frac{1}{2}  \frac{ R_{a}^5  \left( - 7  R_{a} + 8 R_a+14  b \right)  b^3  e^{ \left(R_{a} / b\right) }   c_{sol}  }
{   \left(12  b^2+ R_{a}^2 \right) \left[ -1+  e^{ \left(R_{a}/b \right)} \right]  -   6 b R_{a} \left[ 1+ e^{ \left(R_{a}/b\right) }   \right]     }    \, .
\end{align}

We set $R_a = 10.788$~km, $b = 1$~km, $c_{irr} = 5 \times 10^{-2}$,
and vary the amplitude of the ratio $c_{sol}/c_{irr}$. More
specifically, we consider a purely irrotational vector field, i.e.
$c_{sol} =0$, and two cases with increasing solenoidal amplitude,
respectively, $c_{sol}/c_{irr} = 0.02$ and $c_{sol}/c_{irr} = 0.2$.
In the left panel of Fig.~\ref{fig:test} we show the radial profile of
the analytical solutions (lines) for $\theta=\pi/4$ compared to the
numerical solutions (symbols).  By increasing the solenoidal part of
$\mvb{F}$ the solution changes significantly. This is an effect of a
non-zero coefficient $b$.  Note that in a realistic model the
  discontinuity of $\Delta \mu$ at the crust/core intereface will be
  balanced by the elastic response of the crust.  To study more in
detail the accuracy of our numerical code we average the error of the
relevant quantity, $\m{\nabla} \Psi$, in the grid. First, we evaluate
the error in each point by
\begin{equation}
\Delta  \equiv \frac{      | \m{\nabla}  \Psi  -  \m{\nabla}  \Psi_{an}  | }{  \textrm{max} | \m{\nabla} \Psi_{an} |  }  \label{eq:err}
\end{equation}
and secondly we average the result in all the grid. Note that in
equation (\ref{eq:err}) we have used the maximum as there are points
where $| \m{\nabla} \Psi_{an} |$ vanishes. The variation of the
averaged error $ \langle \Delta \rangle $ with the grid
resolution is shown in the right panel of Fig. \ref{fig:test} for the
three cases with increasing solenoidal component (see legend).  For
the resolution used in this work, 360x360 points, the averaged error
is less than $0.03\%$ for the purely irrotational case, and increases
with the presence of the solenoidal part to $0.1\%$ when
$c_{sol}/c_{irr} = 0.2$.

\label{lastpage}

\end{document}